\documentclass[twocolumn]{aastex701}
\usepackage[utf8]{inputenc}
\usepackage[T1]{fontenc}  
\usepackage{booktabs}  
\usepackage{tabularx}
\usepackage{xcolor}

\usepackage{amsmath}
\usepackage{lipsum} 

\begin{document}

\title{Modeling and Analysis of Main-Belt Asteroidal Dust Flux and Velocity Distribution at Inner Planets}

\author[0009-0006-0630-5536]{Aanchal Sahu}
\affiliation{Planetary Sciences Division, Physical Research Laboratory, Navrangpura, Ahmedabad 380009, India}
\affiliation{Department of Physics, Indian Institute of Technology Gandhinagar, Palaj, 382355, India}
\email[show]{aanchalsahu@prl.res.in}

\author[orcid=0000-0002-0278-5919]{Jayesh Pabari}
\affiliation{Planetary Sciences Division, Physical Research Laboratory, Navrangpura, Ahmedabad 380009, India}
\email{jayesh@prl.res.in}   

\begin{abstract}
Interplanetary dust in the inner solar system originates from multiple sources, including short-period comets and main-belt asteroids. In this work, we focus specifically on the dynamical evolution of asteroid-derived dust using N-body simulations that incorporates solar gravity, planetary perturbations, radiation pressure, Poynting-Robertson drag and solar wind forces. We quantify dust fluxes for Mars Venus and Mercury across an important mass range, which are essential inputs for ablation process on Mars/Venus and for contributing in the impact process on Mercury. We have also derived impact velocity distributions and compared with existing literature. In addition, we examine how close-encounter velocities depend on the orbital elements linking dust energetics directly to the orbital architecture of the dust population. Our results show that the calibrated asteroidal flux agrees excellently with the scaled Grün model for Mars ($0.04$ orders of magnitude offset) and Venus ($0.09$ orders), and with the Müller (2002) model for Mercury ($0.04$ orders). The velocity distributions reveal a decoupling between flux and impact velocity: low-eccentricity grains dominate the flux, while high-eccentricity grains control the high-velocity tail. These findings have direct implications covering: (i) For atmosphere-less bodies like Mercury, the high-velocity tail affects impact processes and exosphere generation; (ii) For Mars and Venus, the flux-dominated low-velocity population determines meteoroid ablation rates and metal layer formation; (iii) Our calibrated fluxes provide inputs for comparison with future observations from different missions and also, for modeling impact-driven surface modification across the inner solar system.

\end{abstract}



\section{Introduction} 
Asteroids and comets have long been considered as two major sources that contribute dust particles to the interplanetary space in inner Solar System \citep{Dermott, GRUN1985244}. The high number density and relative velocities of asteroids indicate frequent occurrence of catastrophic collisions in the main asteroid belt. This was supported by observations of asteroid families—groups with similar orbital parameters that represent fragments of disrupted parent bodies \citep{1918AJ.....31..185H}. Recent work has improved  understanding of the collisional evolution of the main asteroid belt and the formation of young asteroid families \citep{NESVORNY2026116768}. Advances in InfraRed (IR) astronomy, particularly from the IRAS and COBE spacecraft, have fundamentally transformed our understanding of the interplanetary dust cloud (zodiacal cloud). Earlier models assumed a smooth, rotationally symmetric structure centered on the Sun. However, IR observations have revealed that the cloud is highly structured and asymmetric \citep{Dermott_Gomes_Durda_Gustafson_Jayaraman_Xu_Nicholson_1992}. 
We now know that the Sun is not located at the center of symmetry of the zodiacal cloud, which instead exhibits a variety of distinct structures. These include narrow dust bands produced by asteroid family collisions, dust trails associated with known comets, and a circumsolar resonant ring formed by particles trapped in the 1:1 mean-motion resonance (MMR) with Earth \citep{2013Icar..226.1550K}. Among these features, dust bands provide some of the most compelling evidence for an asteroidal contribution to the zodiacal cloud. First identified in IRAS observations as faint, symmetric enhancements at 12 and 25~$\mu$m, these bands are interpreted as swarms of particles sharing similar orbital elements, originating from collisional debris within specific asteroid families. This link between infrared structure and asteroidal source populations provides a foundational basis for modeling the asteroidal component of the impactor flux on terrestrial planets, which is the focus of this investigation \citep{1990Icar...85..267S}. A typical asteroidal dust particle is created by the breakup of a larger parent body, which itself may be a fragment of an earlier collision. Over time, the particle can also become a parent to smaller fragments through further collisions, forming what is known as a collisional cascade. The in-situ and remote sensing observations have provided critical constraints on size distribution of IDP near Earth. Results from the Long Duration Exposure Facility (LDEF) experiments suggest that dominant particles accreted by Earth are typically on the order of 200~$\mu$m in diameter \citep{doi:10.1126/science.262.5133.550}. Similar conclusions were drawn by \citet{GRUN1985244}, based on evidence from lunar microcrater records and micrometeoroid detectors aboard spacecraft. More recent studies using radar and optical meteor observations \citep{Pokornyarticle}, sub-millimetre observations of zodiacal emission from Planck \citep{Ade2014A&A...571A..14P}, and in situ measurements in the inner zodiacal cloud \citep{Szalay2024PSJ.....5..266S} further constrain the size–frequency distribution of IDPs across different mass ranges. These particles are not primordial but are continuously replenished from ongoing asteroidal and cometary activity. Their estimated lifetimes ($\sim10^6$ years) imply that the zodiacal cloud is a dynamic, evolving structure maintained by steady input from these sources consistent with recent models highlighting the role of long-lived grains in shaping the cloud structure \citep{2024PSJ.....5...82P}. \citet{1978A&A....67..381R}, using the same spatial density distribution but deriving scattering efficiencies from Mie theory, reached a similar conclusion on dominant particle sizes. They found that particles with $s < 1\,\mu$m contribute only about 2\% to the zodiacal light, while particles with $s > 10\,\mu$m account for nearly 80\% of the total brightness. This indicates that particles in the range $10\,\mu{\rm m} \lesssim s \lesssim 100\,\mu{\rm m}$ (corresponding to masses of $\sim10^{-8}$--$10^{-5}$ g) dominate the zodiacal cloud in terms of surface brightness and are therefore most relevant for studies of IDP fluxes (considered herein), and impact velocities at the terrestrial planets.

Dust grains originating from asteroid belt gradually evolve due to the combined influence of non-gravitational forces and gravitational pull of the Sun. Radiation pressure can exceed solar gravity for sufficiently small particles when the radiation-to-gravity ratio  $\beta > 0.5$, resulting in their ejection from the Solar System. The value of $\beta$ depends on particle size, composition, and density, and may decrease again for submicron grains \citep{WILCK1996493}. Recent in situ and remote observations have further characterized the properties and variability of $\beta$-meteoroids in the inner heliosphere (e.g., \citealt{2020ApJ...892..115M, 2021PSJ.....2..185S, 2024ApJ...972...24S}). These particles, detected by dust detectors on Pioneers 8 and 9, are classified as $\beta$-meteoroids \citep{ZOOK1975183}. Conversely, gravity prevails over radiation pressure for more massive particles, which maintain nearly circular orbits. This population, observed by Helios 1 and 2 spacecraft, constitutes the $\alpha$-meteoroids \citep{1980P&SS...28..333G}. Specifically, particles smaller than 1 cm  experience significant perturbations from the Poynting-Robertson drag (hereafter PR Drag) and solar wind drag, causing them to spiral inward toward the Sun over timescales that vary according to their size and composition \citep{Borin2017}. Several studies have explored the dynamical evolution of IDPs using numerical models, highlighting the roles of radiation forces, planetary perturbations, and source-dependent orbital evolution \citep{2010ApJ...713..816N, 2011ApJ...743..129N, 2018ApJ...863...31P}. The orbital evolution of dust grains under the combined influence of solar radiation and gravitational forces is well established \citep{1950ApJ...111..134W, BURNS19791}. However, when planetary gravitational perturbations are additionally considered, the dynamical behavior becomes significantly more complex \citep{GUSTAFSON1987568, GUSTAFSON1986280}. Beyond simple gravitational scattering encounters with planets, dust particles can become temporarily captured in MMRs, substantially altering their orbital evolution \citep{1992Icar...97...70J}. There is also a drag force arising from the interaction between dust particles and solar-wind protons. On average, this solar-wind drag amounts to about $30\%$ of the PR drag \citep{1990pihl.book..207L, KORTENKAMP1998469}. Together, the P-R drag and solar-wind drag dissipate orbital energy and angular momentum, causing a gradual decay in the semi-major axis and eccentricity of dust particle orbits. The intricate coupling between planetary perturbations and dissipative non-gravitational forces leads to rich dynamical behavior that governs the final fate of interplanetary dust particles (IDPs).

As planets and their natural satellites travel along their orbital paths, they are persistently exposed to IDP fluxes. When these particles encounter a planetary atmosphere, their high entry velocities result in rapid deceleration and intense heating, leading to ablation. Gravitational acceleration and collisional interactions with atmospheric constituents causes significant thermal energy dissipation, resulting in the removal of surface material through melting and vaporization. This atmospheric processing of micrometeoroids not only alters their physical structure but also introduces exogenous species into the atmosphere. These deposited elements can modify the atmospheric composition and influence the vertical distribution of plasma density in the ionosphere, as demonstrated in previous studies \citep{Molina2008article, PABARI2023105617}. \citet{doi:10.1126/science.1117755} reported evidence for a third ionospheric layer on Mars at an altitude of approx. $85\,\mathrm{km}$, attributed to meteoroid impacts. This layer was later confirmed by \citet{withers} using MGS/RS observations. The Martian atmosphere, being both thinner and supported by a lower gravitational potential, allows a greater fraction of incoming particles to survive atmospheric entry with minimal alteration. This makes Mars a compelling environment for studying nearly pristine micrometeoritic material that has not undergone significant thermal degradation. Recent modeling studies have further quantified the total IDP mass fluxes and resulting meteoric ablation rates in the atmospheres of Earth, Venus, and Mars, including their seasonal and latitudinal variability \citep{CARRILLOSANCHEZ2020113395, 2022PSJ.....3..239C}. 

Consequently, assuming a uniform and steady flux of dust grains from the main belt into the inner Solar System may not be accurate. A comprehensive numerical approach is necessary to accurately model how the dust population evolves as it moves toward the Sun. Understanding the mass influx of IDP into planetary atmospheres is essential for assessing their atmospheric and surface impacts. Accurate dust models are needed to characterize the production, transport, and evolution of dust grains from their sources to planetary atmospheres. \citet{PABARI2023105617} suggested a new interplanetary dust flux model for Venus based on available observations. \citet{HIRAI201487} reported the IDP flux near Venus to be on the order of $10^{-4}$–$10^{-5}~\mathrm{m^{-2}\,s^{-1}}$ based on ALADDIN observations. Additional measurements have also been obtained from the Helios and Galileo instruments \citep{33057761b02e4130b09a7d84297da8c2}. Evidence for dust enhancement just outside Venus' orbit was first identified from \textit{Helios} observations \citep{Leinert2007refId0}. Subsequent \textit{STEREO} measurements revealed the presence of a circumsolar dust ring near Venus, characterized as a two-step ring structure \citep{doi:10.1126/science.1243194, JONES2017172}. Recent studies have identified dust enhancements associated with planetary orbits, including the Venus co-orbital dust ring \citep{Pokorný_2019} and a circumsolar dust ring near Mercury’s orbit detected observationally and modeled dynamically \citep{Stenborg_2018, Pokorný_2023}. More recently, the WISPR (Wide-field Imager for Solar Probe) instrument aboard the Parker Solar Probe detected an excess brightness of $\sim 1\%$ relative to the background zodiacal light, attributed to an increased dust number density near the orbit of Venus. The inferred average density enhancement is approx. $10\%$ \citep{stenborg2021pristine}. These detections suggest that dust in the vicinity of Venus' orbit may partly consist of ring particles gravitationally influenced or temporarily captured by the planet.

In the case of Mars, Phobos and Deimos are continually impacted by interplanetary meteoroids, and, owing to their weak gravitational fields, numerous ejecta particles can escape from their surfaces. These escaped particles disperse along the orbits of their parent satellites due to their relative motion, ultimately forming diffuse dust rings around Mars \citep{ISHIMOTO1996153}. Dust particles in the mass range $\sim 10^{-11}$--$10^{-7}$ kg (size $\sim 10$--200 $\mu$m), potentially released from the Martian moons, can initially form a circum-Martian ring \citep{PABARI20171}. The Juno spacecraft detected a local enhancement in the IDP flux near Mars’s orbit \citep{https://doi.org/10.1029/2020JE006509}. The Juno observations were explained by \citet{10.1093/mnras/stad1045} through modeling, which showed that Phobos and Deimos are the local sources of dust near Mars. Comparison of Juno observations with dynamical meteoroid models show differences in impact rates with expected zodiacal dust or Mars-originating particles \citep{Pokorný_2022}. The Mars Dust Counter (MDC) instrument aboard the Nozomi mission was designed to investigate high-altitude dust phenomena at Mars. During its cruise phase, the MDC detected $\sim$100 particles, with spectral analysis confirming several of these as interplanetary dust particles \citep{2002M&PSA..37R.126S}. The fate of incoming particles depends critically on three key parameters: mass, velocity, and entry angle. Depending on these factors, dust grains may either completely ablate during atmospheric entry or survive as unmelted micrometeorites if they avoid reaching their melting temperature. Using the measured mass influx at Earth and estimates of the Mars-to-Earth flux ratio, the continuous, planet-wide meteoritic mass influx on Mars has been inferred to lie between $2.7\times10^{3}$ and $5.9\times10^{4}\,\mathrm{t\,yr^{-1}}$ \citet{1990JGR....9514497F}. Recent modeling suggests that the meteoroid mass input at Mars varies seasonally between about 1.5 and 2.3 tons per sol \citep{2022PSJ.....3..239C}. MAVEN observations provide additional constraints on the dust environment at Mars. The LPW instrument detected micron-sized grain impacts, corresponding to a mass influx of only $\sim 0.001$-$0.1\,\mathrm{kg\,s^{-1}}$, indicating that the detected particles represent just a small fraction of the total interplanetary dust input \citep{Andersson2015Sci...350.0398A}. The \citet{PABARI20171} study used MAVEN observations and suggested dust flux model for Mars. They showed that the dust at high altitudes of Mars is expected to be interplanetary in nature. The Mars Orbiter Dust Experiment (MODEX) has been proposed as a future payload to characterize the circum-Martian dust environment \citet{Pabari2016MARSOD}. The upcoming Martian Moons eXploration (MMX) mission will carry the Circum-Martian Dust Monitor (CMDM), designed for in situ measurements of dust around Mars and its moons \citep{KOBAYASHI201841}.

In this study, we investigate long-term dynamical evolution of dust grains originating from main-belt asteroids, as they migrate toward the inner solar system. Using a numerical model that accounts for both gravitational and non-gravitational forces, we track the orbital trajectories of these particles. Through simulations, we quantify the resulting dust flux, impact velocity distribution upon reaching planets and encounter velocity dependence on orbital elements. The remainder of this paper is organized as follows. Section~\ref{sec:style} describes the dynamical evolution model adopted to simulate the long-term behavior of asteroidal dust, including both gravitational and non-gravitational forces implemented in the numerical integrator. Section~\ref{sec:flux_calculation} outlines the methodology used to derive the dust impact flux on planetary bodies, together with the calibration procedure. The resulting velocity distributions of dust grains impacting the planets are presented in Section~\ref{sec:vel}. Section~\ref{sec:initial} details the initial conditions assumed for the asteroidal dust population. The simulation results are presented and discussed in Section~\ref{sec:results}, including the dependence of the radiation-pressure parameter $\beta$ on dust size, the flux and velocity distributions for multiple grain sizes, and a comparative analysis of dust--planet encounter velocities as functions of orbital eccentricity, inclination, and longitude of perihelion for the inner planets. Finally, Section~\ref{sec:conclusion} summarizes the main findings of this study and discusses their implications for future missions, as well as for the dust environment of the inner solar system.

\section{Methodology}
In this section, we outline the computational framework and analytical procedures used to model the evolution of IDP and to estimate the corresponding impact fluxes at the inner planets. Our methodology combines long-term N–body integrations that include both gravitational and non-gravitational forces, statistical treatment of close encounters, and calibration of simulated fluxes against observational constraints.

\subsection{Dynamical Evolution Model} \label{sec:style}
To estimate the meteoritic flux at the heliocentric distance of different planets, we used the dynamical evolution model of dust particles of Marzari \& Vanzani \citep{1994A&A...283..275M}. It numerically integrates an (N+1)+ M body problem (Sun + N planets + M bodies with negligible mass) with the high-precision integrator RA15 version of the RADAU integrator by \citet{Everhart1985} under the effect of gravitational and non- gravitational forces. For small dust particles orbiting the Sun, one of the primary non-gravitational forces influencing their long-term dynamics is solar radiation which manifests as two distinct effects: a radial component referred to as radiation pressure, and a tangential component known as P-R drag {\citep{BURNS19791}. While radiation pressure can expel extremely small particles from the solar system, P–R drag becomes dominant for micron-sized particles leading to a continuous decrease in the semi-major axis, with a more rapid contraction of the aphelion compared to the perihelion. As a result, the orbital eccentricity diminishes over time, and the particle spirals inward toward the Sun. The inclination of the orbit remains largely unaffected by this process, as the drag force does not exert a component perpendicular to the orbital plane. The model accounts for perturbative forces, including radiation pressure, solar wind drag, and P-R drag, along with the gravitational influence of all planets in the solar system  \citep{2016A&A...588C...3B, 2009A&A...503..259B}. Collisions between dust particles which can shape the zodiacal cloud \citep{Rigley10.1093/mnras/stab3482} are omitted here, as our mass range ($10^{-9}$--$10^{-4}$~g) corresponds to grain radii mostly below 125 $\mu m$, for which P-R drag exceeds collisional disruption rates \citep{2024PSJ.....5...82P}. A full treatment of collisions would require additional assumptions \citep{Stark2009ApJ...707..543S} and is beyond the scope of this work.

Employing the methodology of \citet{1994A&A...283..275M}, the gravitational component is formulated as
\begin{equation}
    \mathbf{F}_{\text{gra}} = \mathbf{F}_{k} + \mathbf{F}_{d} + \mathbf{F}_{\text{ind}},
\end{equation}
where \( \mathbf{F}_{k} \) is the Keplerian force, \( \mathbf{F}_{d} \) is the direct force, which includes the interaction between planets and dust particles, and \( \mathbf{F}_{\text{ind}} \) is the indirect force representing the interaction between the central body and dust particles. Equation (1) can be rewritten as
\begin{equation}
    \mathbf{F}_{\text{gra}} = \frac{Gm(M_{\odot} + m) \mathbf{r}_{\odot}}{r_{\odot}^3} + \sum_{j=1}^{N} \frac{Gm m_j \mathbf{r}_j}{r_j^3} 
    + \sum_{j=1}^{N} \frac{Gm m_j \mathbf{r}_{\odot, j}}{r_{\odot, j}^3},
\end{equation}
where $\mathbf{r}_{\odot}$ is the distance between the Sun and dust particles, $\mathbf{r}_j$ is the distance between planets and dust particles, $m$ is the mass of dust particles, and $N$ is the number of planets.
The non-gravitational force consists of two terms: the radiation force $\mathbf{F}_{\text{rad}}$ and the force due to the (corpuscular) solar wind drag $\mathbf{F}_{\text{wnd}}$ :

\begin{align}
    \mathbf{F}_{\text{ngra}} &= \mathbf{F}_{\text{rad}} + \mathbf{F}_{\text{wnd}},
\end{align}

with

\begin{equation}
    \mathbf{F}_{\text{rad}} = \frac{S}{c} \left(1 - \frac{\dot{r}}{c}\right) A Q_{\text{pr}} \hat{\mathbf{p}} = f_r \hat{\mathbf{p}}, 
\end{equation}
\\
the resistive force on the dust grain due to its interaction with the solar radiation

\begin{equation}
    \mathbf{F}_{\text{wnd}} = \frac{\eta_j u^2}{2} A C_{D,j} \hat{\mathbf{u}} = f_w \hat{\mathbf{u}}.
\end{equation}
\\
the corpuscular counterpart of $\mathbf{F}_{\text{rad}}$ with a self explanatory notation of $f_r$ and $f_w$. In Equation (4) and (5), $\hat{\mathbf{p}} = \frac{\mathbf{c} - \mathbf{v}}{c}$, where $\mathbf{c}$ is the velocity of light (in the anti-solar direction), and $\mathbf{v}$ is the orbital velocity of the dust particle. Additionally, $\hat{\mathbf{u}} = \frac{\mathbf{u}}{|\mathbf{u}|}$, where $\mathbf{u} = \mathbf{w} - \mathbf{v}$, where $\mathbf{w}$ is the solar wind flow bulk velocity in the average phase \citep{1994A&A...283..275M, 1982A&A...107...97M}; the spatial mass density of the $j$-th component of the solar wind is given by $\eta_j = n_j m_j$ with mass $m_j$ and number density $n_j$. $A$ is the geometrical cross-section of the grain, $Q_{\text{pr}}$ is the dimensionless radiation-pressure coefficient averaged over the solar spectrum, $C_{D,j}$ is the drag coefficient due to the $j$-component of the wind flow, and $S$ is the solar radiation flux density at heliocentric distance $r$ with $S = S_0 \left( \frac{r_0}{r} \right)^2$; \( w_0 \approx 4 \times 10^7 \, \text{cm/s} \) for \( w \) at 1 AU; and \( \eta_{p,0} + \eta_{\alpha,0} \approx 1.2 \eta_{p,0} \). This solar wind drag force is taken about 30\% of the PR drag in our model \citep{1990pihl.book..207L, KORTENKAMP1998469}. The standard computation of $Q_{\mathrm{pr}}$ is based on Mie theory, including detailed optical constants of the grain material and the solar energy spectrum. For the drag coefficient $C_{D,j}$, we adopt the results of  \citet{1982A&A...107...97M}, who account for the velocity dispersion of wind particles and include the contribution to the solar-wind drag arising from the momentum carried by sputtered molecules.

The efficiency of the radiation and corpuscular resistive forces can be expressed by defining their ratio with respect to solar gravity in the following manner:

\begin{equation}
     \beta_r = \frac{f_r}{f_g} \left[\frac{c}{c - \dot{r}}\right] =\left( \frac{S A Q_{\text{pr}}}{c} \right) \left( \frac{GM_{\odot} m}{r^2}\right)^{-1}
\end{equation}
and
\begin{equation}
     \beta_w = \frac{f_w}{f_g}\left[ \frac{w}{|\mathbf{w} - \mathbf{v}|} \right] = \left( \frac{f_{\omega 0} \psi}{\kappa} \right)\left( \frac{GM_{\odot} m}{r^2}\right)^{-1}
\end{equation}
for $\kappa = \frac{u}{w}$ and $\psi = \frac{f_w}{f_{w0}}$, where $f_{w0}$ is obtained from $f_w$ when neglecting the velocity dispersion of wind particles and the contribution of momentum carried away by sputtered molecules to $F_{\text{wnd}}$ \citep{BURNS19791}. \\

Taking the reference-distance $r_0$ to be equal to 1 AU and assuming a dust particle of spherical shape of radius $s$, one obtains

\begin{equation}
    \beta_r = \frac{3 S_0 r_0^2}{4 GM_{\odot} c} \frac{Q_{\text{pr}}}{\rho s} = 5.74 \times 10^{-5} \frac{Q_{\text{pr}}}{\rho s},
\end{equation}
and
\begin{equation}
    \beta_w = \frac{3 (\eta_{p,0} + \eta_{\alpha,0}) r_0^2 w_0^2}{4 GM_{\odot}} \frac{\psi \kappa}{\rho s} \simeq 3.27 \times 10^{-8} \frac{\psi \kappa}{\rho s},
\end{equation}
\\    
where $\rho$ is the mass density of the dust particle in CGS units. Finally, the relative importance of radiation forces and corpuscular forces can be quantified through the parameter
\begin{equation}
\gamma = \frac{\beta_{w}}{\beta_{r}}
\simeq 5.7 \times 10^{-4} \,
\frac{\psi \kappa}{Q_{\mathrm{pr}}},
\end{equation}
In terms of $\beta_{r}$ and  $\gamma$, the combined contribution of radiation and solar resistive forces can be written as
\begin{multline}
\mathbf{F}_{\mathrm{ngra}} =
\beta_{r} f_{g} 
\left[
(1 + \gamma \cos\phi)\,\hat{\mathbf{r}}
\mp \gamma (\sin\phi) \,\hat{\boldsymbol{\vartheta}}
\right]
\\
-
\beta_{r} f_{g}
\left[
\frac{\dot{r}}{c}\,(2 + \frac{\gamma c}{w})\,\hat{\mathbf{r}}
+
\frac{r\dot{\vartheta}}{c}\,(1 + \frac{\gamma c}{w})\,\hat{\boldsymbol{\vartheta}}
\right],
\end{multline}
where the terms independent of the particle velocity are separated  from those proportional to the grain velocity. Here, $\hat{\boldsymbol{\vartheta}}$ is the transverse unit vector in the orbital plane, positive in the direction of motion, and $\phi = \arccos(\mathbf{w}\cdot\hat{\mathbf{r}})/w$ is the angle between the average solar–wind flow direction and the grain velocity. The first bracketed term in Eq.~(11) represents the sum of the radiation–pressure force, $\beta_{r} f_{g}\,\hat{\mathbf{r}}$, and the corpuscular–pressure force, $\beta_{w} f_{g}\,\hat{\mathbf{w}} = f_{w}\,\mathbf{w}/u$, resolved into radial and transverse components. The second bracketed term corresponds to the classical PR drag combined with the solar–wind drag.
Lorentz forces arising from the interaction of charged dust grains with the interplanetary magnetic field are considered negligible for the particle sizes examined in this study (radii s = 5 -- 200$\mu m$), as their charge-to-mass ratios are expected to be sufficiently small \citep{Mukai:1984:MSD, GUSTAFSON1986280}.

To numerically integrate the orbits of dust particles under the influence of aforementioned forces, we employ the Everhart RA15 integrator \citep{Everhart1974, Everhart1985}, a high-order, variable-step, implicit Runge-Kutta method specifically designed for celestial mechanics problems. The RA15 scheme is well-suited for long-term integrations due to its high accuracy and numerical stability, especially in handling close approaches and stiff dynamical regimes. The formulation builds upon standard principles of orbital dynamics and perturbation theory (e.g., \citealt{roy2004orbital}; \citealt{Fitzpatrick_2012}), while adopting the numerical implementation of Everhart’s method for efficient long-term evolution.

\subsection{Impact Flux Calculation} \label{sec:flux_calculation}

\begin{figure*}[ht]
    \centering
    \includegraphics[width=0.7\textwidth]{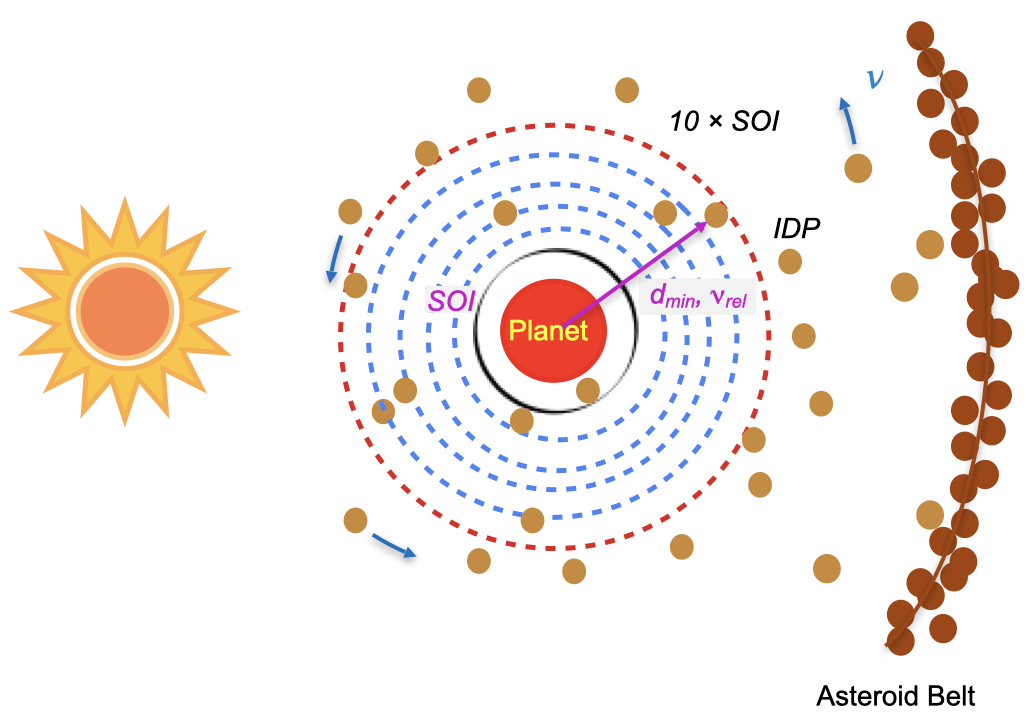}
    \caption{Schematic representation of the geometry adopted to estimate the dust impact flux from close encounters. Dust particles originating from the asteroid belt are shown as small incoming trajectories migrating toward the Sun. The planet is located at its heliocentric distance $a_{\mathrm{planet}}$; the solid black circle denotes the planetary Hill sphere, while the red dashed circle marks the adopted sphere of influence (SOI): $10\,R_{\mathrm{H}}$ for inner planets. Blue dotted rings illustrate the concentric radial bins used to sample encounter distances (figure not to scale).}
    \label{fig:flux_img}
\end{figure*}
Classical collisional probability methods (e.g., \citep{1951PRIA...54..165O, Wetherill1967CollisionsIT, POKORNY2013682}) provide analytical estimates of impact rates. We instead compute fluxes directly from N-body integrations by tracking planet-crossing particles, which accounts for gravitational focusing and time-dependent orbital evolution. The numerical integrations employed in this work are based on the
$N$-body code originally developed by \citet{MARZARI1996192} for the dynamical evolution of IDPs. While the orbital integration framework follows this established approach, the procedure adopted here to estimate planetary dust impact fluxes is entirely implemented within the present work. In particular, we introduce a dedicated post--processing methodology based on the systematic analysis of close encounters between dust particles and planets, allowing us to statistically infer impact fluxes even in regimes where direct impacts are rare. Our numerical simulations track the orbital evolution of dust particles until their semimajor axes decay well inside the orbit of the planet under consideration. To quantify the impact flux, we employ a statistical approach due to the limited number of direct impacts in our simulations. As dust particles approach the planet, we monitor their trajectories and identify close encounters occurring within a prescribed sphere of influence. The planet is assumed to be located at its heliocentric distance $a_{\mathrm{planet}}$.
The characteristic scale of the encounter region is given by the Hill radius,
\begin{equation} 
R_{\mathrm{H}} = a_{\mathrm{planet}} \left( \frac{m_{\mathrm{planet}}}{3 M_{\odot}} \right)^{1/3},
\end{equation}
where $m_{\mathrm{planet}}$ is the planetary mass and $M_{\odot}$ is the solar mass. The Hill radius  \citep{Hill0eb94dbb-e87d-35f4-a82b-f98b26a6e529} provides an approximate boundary separating regions where the planet’s gravity dominates the local dynamics from those dominated by the Sun. For inner planets, we adopt a sphere of influence (SOI) with radius $10\,R_{\mathrm{H}}$. When a dust particle enters this region, we record the minimum grain--planet distance and the relative velocity at the closest approach. These close encounter data form the basis for our statistical analysis. The encounter region is divided into a series of concentric radial bins centered on the planet, and encounter distances of dust grains entering this region are accumulated within the corresponding bins. These binned encounter statistics are used to construct the spatial probability distribution of close encounters which is fitted using a parabolic function as $P_0 R^2$, where $R$ is the radial distance from the planet’s center and $P_0$ is a fitting constant. This quadratic dependence arises as for a uniform 3-D particle distribution, the encounter cross section and encounter probability density scales as $R^2$ in spherical coordinates \citep{MARZARI1996192}.

The fitting procedure employs a weighted least-squares approach in which each radial bin is weighted by $\sqrt{N_i}$, where $N_i$ is the number of encounters in the $i$-th bin. This weighting scheme naturally reflects the Poisson statistics of the encounter counts. The derived scaling parameter $P_0$ is then used to calculate the fractional number of impacts $n_M$ at the planetary surface by setting $R$ as the planetary radius. Gravitational focusing is also included to account for the enhanced impact probability due to the planet’s gravitational attraction. A population of 1000 simulated particles of each size provides a sufficiently robust representation of the dynamical behaviour of real dust grains during their approach to the planet. From the computed number of impacts $n_M$, the corresponding flux $g_M$ is obtained by dividing $n_M$ by the effective time interval $\Delta T$ during which the impacts occur. In our simulation, we define an effective time interval $\Delta T$ from the close encounter distribution with time during which the encounter rate remains approximately steady. The temporal distribution of close encounters typically exhibits three distinct phases: an initial rise as the first dust grains drift into the planet’s vicinity, a plateau where the encounter rate is nearly constant, and a final decline as the population of inward-migrating grains diminishes. Following \citet{Borin2017}, we identify $\Delta T$ as the duration of this steady phase, during which the flux is statistically representative of sustained dust delivery.
The impact flux $g_M$ on the planet is then estimated by:
\begin{equation}
    g_M = \frac{n_M'}{\Delta T},
\end{equation}
where $n_M'$ is rescaled to represent the fractional number of impacts expected over the effective interval $\Delta T$. The gravitational attraction of planetary bodies significantly enhances their effective capture cross-sections for incoming dust particles. This gravitational focusing effect amplifies the dust flux by a factor 
\citep{MOLINACUBEROS2001143, Molina2008article}:

\begin{equation}
G = 1 + \left( \frac{v_{\mathrm{esc}}}{v_{\infty}} \right)^2
\end{equation}

\noindent where $v_{\mathrm{esc}}$ represents the escape velocity at the 
planetary surface and $v_{\infty}$ is the relative velocity at infinity 
between the particle and the planet.

\subsubsection{Flux Calibration}
To improve the robustness of our flux estimation, we further calibrate the dust density in our simulation using observational data from Earth. We record encounters of dust grains with Earth and compute the corresponding flux $g_E(s)$ for each particle size $s$. This allows us to define a set of calibration coefficients:
\begin{equation}
    C(s) = \frac{g_M(s)}{g_E(s)}.
\end{equation}
The coefficient $C(s)$ quantifies the relative variation in the flux of dust grains of radius $s$ between Earth and the planet of interest (e.g., Mars). It implicitly accounts for dynamical effects such as mean-motion resonances, close planetary encounters, and differences in inward migration rates, following the methodology of \citet{2009A&A...503..259B}, where analogous coefficients are used to scale the meteoroid flux from Earth to other planets. To convert our simulated fluxes into absolute physical values,  we calibrate against three independent Earth-based datasets: LISA Pathfinder \citep{Thorpe_2019}, the asteroidal flux estimate of \citet{CARRILLOSANCHEZ2020113395}, and the empirically determined meteoroid flux at Earth derived by \citet{2012ApJ...749L..40C}. For each mass bin, we compute the average of the three fluxes to obtain the final Earth calibration curve. The flux on each target planet is then obtained by multiplying this Earth calibration curve by the coefficients C(s)
with a smooth cubic spline drawn through the discrete mass-bin values for visual representation. This procedure provides an observationally anchored estimate of the absolute dust flux fully accounting for both the dynamical evolution tracked in our simulation and the latest empirical constraints.

\subsection{Impact Velocity Distribution} \label{sec:vel}
The impact velocity distribution is derived from close encounter data, corrected for gravitational focusing. Our method allows direct extraction of the impact velocity distribution from the simulation outputs. These velocities naturally reflect the combined effects of dynamical processes such as orbital evolution under PR drag, resonance-driven migration, and gravitational scattering by planets. The impact velocity $v_{\text{imp}}$ of a dust particle is calculated using the following relation:
\begin{equation}
    v_{\text{imp}} = \sqrt{v_{\text{esc}}^2 + v_{ce}^2}
\end{equation}
where $v_{\text{esc}}$ is the escape velocity of the target planet and $v_{ce}$  represents the relative dust–planet velocity obtained from the close-encounter outputs of the N-body simulations. In the gravitational focusing formalism (Eq. 16), this quantity plays the role of the asymptotic velocity $v_{\infty}$. While $v_{\infty}$ is formally defined in an ideal two-body framework, in our case the encounter velocity derived from the full N-body geometry provides a more realistic estimate of the effective incoming speed of particles at planetary distances. The resulting velocity distributions reflect the particles’ orbital histories and energetics, and their dependence on eccentricity, inclination, and longitude of perihelion is examined for inner planets.

\section{Initial Conditions in the Simulation} \label{sec:initial}
\begin{figure*}[ht]
    \centering
    \includegraphics[width=0.98\textwidth]{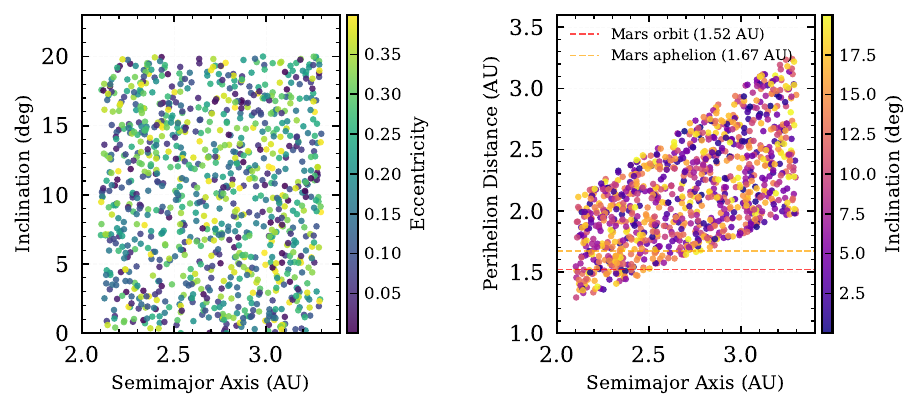}
    \caption{Initial distribution of the dust particles in orbital element space. 
    The left panel shows the inclination as a function of semimajor axis, with color indicating eccentricity. 
    The right panel displays the perihelion distance as a function of semimajor axis, with color indicating inclination. 
    The dashed horizontal lines mark the mean orbital distance and aphelion distance of Mars. 
    Together, these panels illustrate the initial dynamical configuration of the dust population used in the simulation.}
    \label{fig:flux_img}
\end{figure*}

\citet{2010ApJ...713..816N} found that JFCs can account for the majority of the observed IRAS emission. In the present work we instead isolate dust originating from the main asteroid belt to quantify its dynamical evolution and resulting impact signatures at the terrestrial planets. While young asteroid families may act as time-dependent dust sources, modeling their detailed collisional evolution is beyond the scope of the present study. We therefore adopt a simplified steady-state representation of the asteroidal component.
We simulated a prograde ring of 1000 asteroidal dust particles, with initial orbital elements following the distributions adopted by \citet{2009A&A...503..259B, Borin2017}. The semimajor axis was randomly sampled in the range $2.1$--$3.3$~AU, corresponding to the main asteroid belt. The initial eccentricity and inclination were drawn from uniform distributions spanning $0.0$--$0.4$ and $0^\circ$--$20^\circ$, respectively. These ranges are consistent with the typical orbital parameters of the asteroid belt \citep{1989aste.conf..316G, Sykes1986, Milani1994, Sykes2004}. The angular orbital elements—the longitude of the ascending node $\Omega$, the argument of perihelion $\omega$, and the mean anomaly $M$—were randomly distributed between $0$ and $2\pi$ for all particles. This choice ensures an initially azimuthally uniform dust ring and avoids introducing any preferred orbital orientations or phase correlations in the particle ensemble. The simulations were carried out for discrete dust particle radii of 
$s = 5$, $10$, $15$, $20$, $25$, $30$, $50$, $100$, $150$, and $200~\mu$m. We simulated 1000 particles for each of the 10 size bins, resulting in a total of 10,000 particles across the full size range to ensure robust sampling of the dynamical evolution and impact distributions. The corresponding particle masses were assigned assuming spherical grains with a bulk density of $\rho = 2.5~\mathrm{g\,cm^{-3}}$, appropriate for particles of asteroidal origin \citep{GRUN1985244}. The selected size range represents the population of grains that contribute most effectively to the steady-state dust flux at the inner planets considered in this study. We treated the particles as spherical, applying Mie theory with a solar-spectrum-averaged radiation pressure coefficient \(Q_{pr} = 0.53\) \citep{1982A&A...107...97M, 1994A&A...283..275M}. The initial orbital elements of the planets were taken from the JPL Horizons ephemerides at a Julian date JD = 2414837.498, which was adopted as the reference epoch for all integrations.

Figure~\ref{fig:flux_img} illustrates the resulting initial orbital configuration of the dust population. The left panel demonstrates that the simulated particles populate a broad inclination distribution up to $20^\circ$ across the full range of semimajor axes, while the color bar highlights the spread in eccentricity. The right panel shows the corresponding perihelion distances, indicating that a fraction of particles initially possess perihelia close to or interior to the orbit of Mars, even before inward migration driven by non-gravitational forces.

To capture all relevant close encounters with the inner planets, the dust particles were numerically integrated over timescales of several million years. The chosen time span allows even the slowest inward-migrating particles, primarily affected by PR drag, to drift from the asteroid belt into the inner solar system. Integrations used a fixed 1-day timestep with adaptive output: daily during close encounters in order to compute the minimum distances with high precision and every 4 years otherwise. The long integration duration was critical for capturing statistically significant close encounters, particularly for low-$\beta_r$ particles that spiral inward more gradually. This setup ensures that our results reflects a comprehensive sampling across the full dynamical range of particle behaviors in the inner solar system.

\section{Results and Discussion} \label{sec:results}
In this section, we present the main outcomes of our numerical simulations and analytical calculations. The results are organized into several interconnected aspects. First, we examine the size-dependent effects of radiation forces on dust grains, as reflected in the radiation pressure parameter $\beta$ and the modified solar gravitational parameter $g_{\rm msb}$ (Section~\ref{sec:beta_gmsb}). Next, we illustrate the dynamical evolution of dust particles, highlighting the mechanisms of resonance trapping, PR drag, and close planetary encounters (Section~\ref{sec:dynamics}). Finally, we analyze the resulting impact fluxes, calibrate the simulated fluxes against observational constraints, and discuss the distributions of impact velocities at the inner planets and also examine how orbital elements control close-encounter velocities. (Sections~\ref{sec:flux_results} and \ref{sec:velocities_results}). 

\subsection{Radiation Pressure and Modified Solar Gravity} \label{sec:beta_gmsb}
\begin{figure*}
    \centering
    \includegraphics[width=0.82\textwidth]{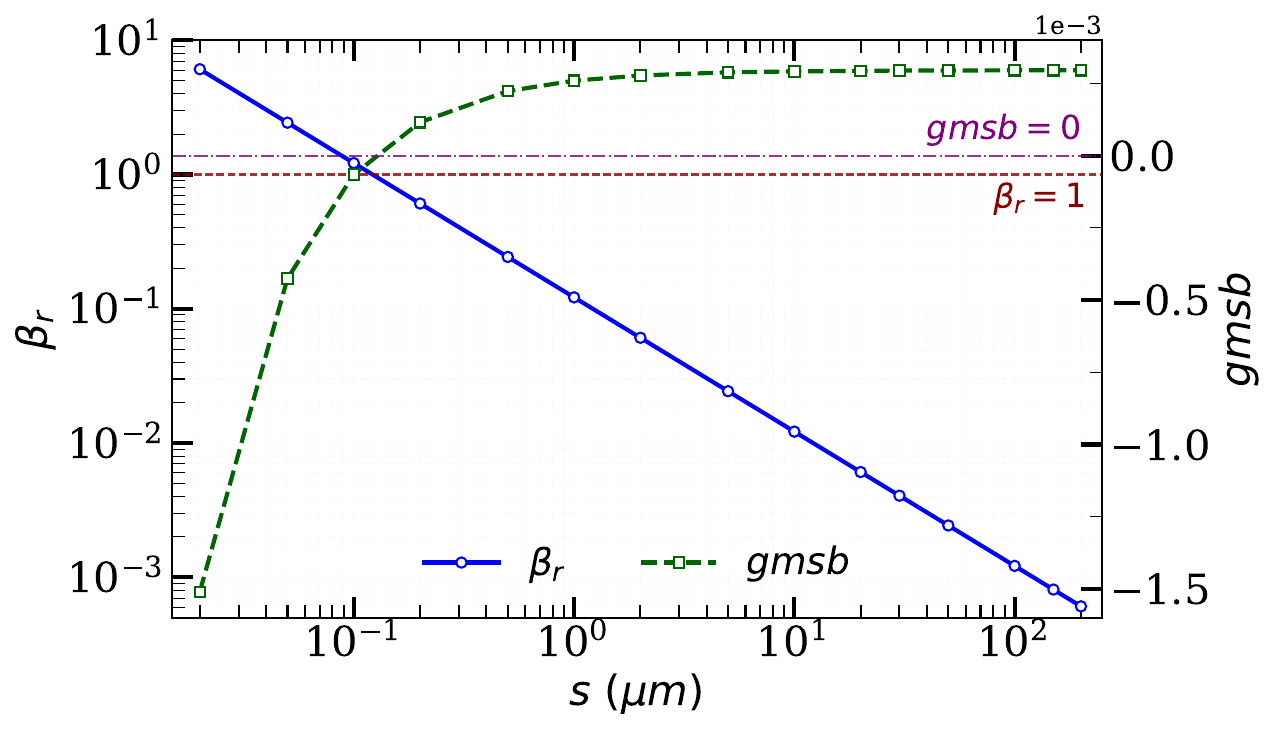}
    \caption{Dependence of radiation pressure parameter $\beta_r$ (blue circles, left axis) and modified gravitational parameter $g_{\text{msb}}$ (green squares, right axis) on particle radius. The dashed line indicates $g_{\text{msb}} = 0$, with the transition from negative to positive values occurring at $s \approx 0.2~\mu$m. Small particles ($s < 0.1~\mu$m) show $\beta_r > 1$, indicating radiation pressure dominance, while large particles ($s > 1~\mu$m) approach asymptotic behavior with negligible radiation pressure effects.}
    \label{fig:beta_gmsb}
\end{figure*}
The radiation pressure parameter $\beta_r$ shown in Fig.~\ref{fig:beta_gmsb} is computed as defined in Section~\ref{sec:style}, and is introduced following \citet{BURNS19791}. The figure illustrates the strong size dependence of $\beta_r$, which decreases monotonically with increasing particle radius as expected from its inverse proportionality to $s$. For sub-micron grains, $\beta_r$ can exceed unity, implying radiation-pressure-dominated dynamics. However, as noted by \citet{WILCK1996493}, $\beta_r$ in this regime is non-monotonic: it peaks near $\sim 0.1$ $\mu$m, then decreases for smaller grains as the particle becomes smaller than the dominant solar wavelengths and experiences the solar spectrum less efficiently. In contrast, micron-sized and larger particles typically have $\beta_r \ll 1$, and remain gravity-dominated. The value $\beta_r = 1$ corresponds to the exact balance between solar radiation pressure and gravitational attraction. However, dust grains released from parent bodies can become dynamically unbound already for $\beta_r \gtrsim 0.5$, depending on the orbital elements and release location along the orbit \citep{2006A&A...455..509K}. For the adopted grain properties ($Q_{\rm pr}$ and bulk density $\rho$), this corresponds to a particle radius $s \simeq 0.1\,\mu$m. Particles smaller than this size experience a net outward acceleration and can become dynamically unbound from the Sun. This behavior directly explains the transition in the modified solar gravitational parameter $gmsb$ shown in the same Fig ~\ref{fig:beta_gmsb}, delineating the boundary between unbound and gravity-dominated dynamical regimes. The corresponding modified solar gravitational parameter $gmsb$ is introduced to account for the reduction of the effective solar gravity due to radiation pressure and solar wind effects. In the numerical model, it is defined as
\begin{equation}
gmsb = g_\odot - \alpha\, c \left( 1 + b_{\rm sub} \cos \phi \right),
\end{equation}
where $g_\odot$ is the nominal solar gravitational parameter, $c$ is the speed of light, $b_{\rm sub}$ accounts for the solar wind contribution, and $\phi$ is the phase angle. The parameter $\alpha$ is a scaling coefficient related to $\beta_r$, defined such that $\alpha c = \beta_r g_\odot$, ensuring that the radiation pressure acceleration is expressed in units consistent with the solar gravitational parameter.

As particle size decreases, the radiation pressure contribution increases according to the expected scaling $\beta_r \propto s^{-1}$, progressively reducing $gmsb$. For particles with $s \lesssim 0.1~\mu$m, corresponding to $\beta_r > 1$, $gmsb$ becomes negative, indicating that the effective solar force is repulsive and such grains are dynamically expelled from the solar system. This regime represents the classical blowout condition, and provides a natural explanation for the depletion of submicron dust in the inner solar system. A transition region is observed for $s \simeq 0.1$--$0.2~\mu$m, where radiation pressure and gravity are comparable and $gmsb$ changes sign. Particles in this size range experience competing forces, and their evolution is therefore particularly sensitive to non-gravitational perturbations. For larger grains ($s \gtrsim 0.2~\mu$m), $gmsb$ rapidly approaches a constant positive value close to the nominal solar gravitational parameter, demonstrating that radiation forces act only as a perturbation to gravity in this regime and that particle motion is essentially Keplerian. The monotonic decrease of $\beta_r$ is clearly reflected in the behavior of $gmsb$, as radiation pressure scales with particle cross-sectional area while gravity scales with mass. The figure therefore clearly identifies the blowout limit near $s \simeq 0.1~\mu$m, highlighting the strong size dependence of dust dynamics in the inner solar system. Particles smaller than this threshold are efficiently removed by radiation pressure, whereas larger particles remain gravitationally bound and dominate the long-term evolution and impact fluxes investigated in this study. This size-dependent force balance creates natural sorting mechanisms in dust populations providing quantitative boundary for dust particle stability against radiation pressure ejection, offering important constraints for models of zodiacal cloud maintenance and the flux of dust impacting terrestrial planets.

\subsection{Dynamical Evolution of IDPs}  \label{sec:dynamics}
\begin{figure}[h]
    \centering
    \includegraphics[width=0.5\textwidth]{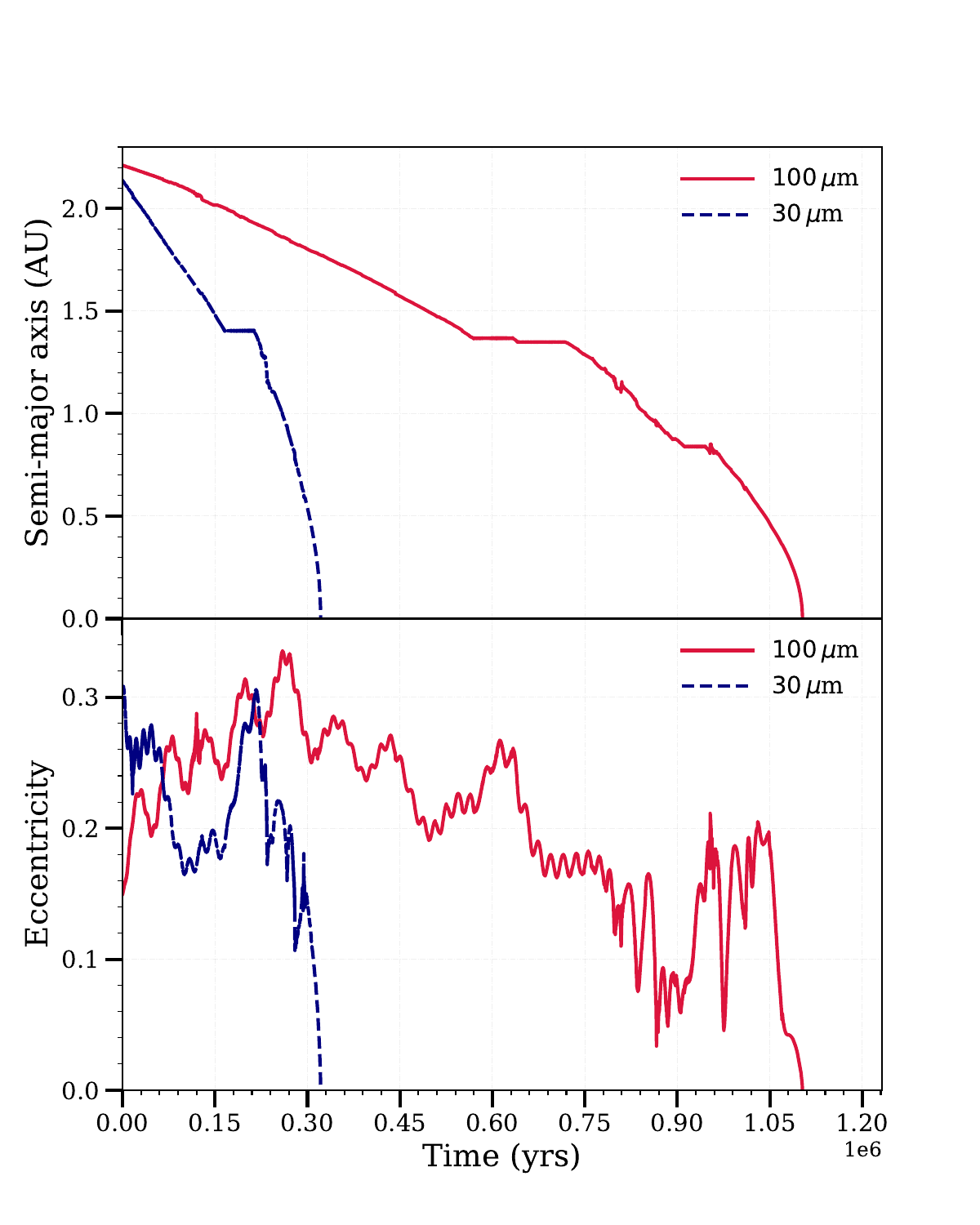}
    \caption{Time evolution of semi-major axis (top) and eccentricity (bottom) for asteroidal dust with radii of $100~\mu$m (solid red line) and $30~\mu$m (dashed blue line). The trajectories correspond to individual particles arbitrarily selected from the simulation to illustrate characteristic dynamical behavior; similar trends are observed across the ensemble.}
    \label{fig:dynamics_doublepanel}
\end{figure}

Before discussing the computed impact fluxes, it is instructive to examine the typical dynamical behavior of dust grains as they migrate through the inner solar system. Figure~\ref{fig:dynamics_doublepanel} illustrates the time evolution of the semi-major axis and eccentricity for representative asteroidal particles of two distinct sizes: $100~\mu$m and $30~\mu$m. These trajectories exemplify the size-dependent mechanisms that regulate the inward drift and orbital excitation of dust grains. \citet{Sommer2020A&A...635A..10S} investigated the formation of resonant dust structures in the inner solar system and found that migrating dust particles can become temporarily trapped in external MMRs with planets, while the efficiency and lifetime of trapping depend strongly on particle size and perturbations from neighbouring planets. Our simulations are broadly consistent with their findings, but we extend their analysis by identifying specific resonances from individual particle trajectories.

Large grains ($100~\mu$m) exhibit significant periods of mean-motion resonance (MMRs) trapping with Earth, during which their eccentricities are periodically excited. In particular, temporary slowing of the inward migration is visible near $\sim1.3$ AU and $\sim1.4$ AU. By evaluating the corresponding mean-motion ratios using Kepler’s third law, these features are found to be consistent with the 2:3 and 3:5 external resonances with Earth, respectively. Similarly, an additional feature near $\sim0.85$ AU is consistent with the vicinity of the 4:3 internal resonance. Additionally, we explicitly verify that the features near 1.3–1.4 AU are not attributable to Mars resonances. This constitutes the first direct, trajectory-level identification of specific MMR locations for the dust population evolving under P-R drag in the inner Solar System. Close planetary encounters occur but typically induce only minor changes in the semi-major axis, so these grains remain in the inner solar system for extended durations. The recurrent resonance captures and the associated eccentricity growth slow the inward migration of these grains and maintain elevated relative velocities during planetary encounters, thereby influencing both the timing and velocity distribution of impacts. Smaller grains ($30~\mu$m), in contrast, are dominated by PR drag and migrate inward more rapidly. While temporary resonance captures may occur, these are brief, and the particles resume their inward drift toward the Sun. Their fast evolution results in shorter dynamical lifetimes in the terrestrial planet region compared with larger grains. These behaviors underscore the size-dependent dynamical sorting of dust particles: larger grains remain longer in resonances and maintain elevated eccentricities, while smaller grains are efficiently transported inward by non-gravitational forces. Understanding these differences is essential, as they strongly influence the impact probabilities and velocities at terrestrial planets, setting the stage for the flux analysis and velocity distribution  presented in the subsequent sections.

\subsection{Flux Estimation on Mars, Venus and Mercury}  \label{sec:flux_results}
\begin{figure*}
    \centering
    \includegraphics[width=0.77\textwidth]{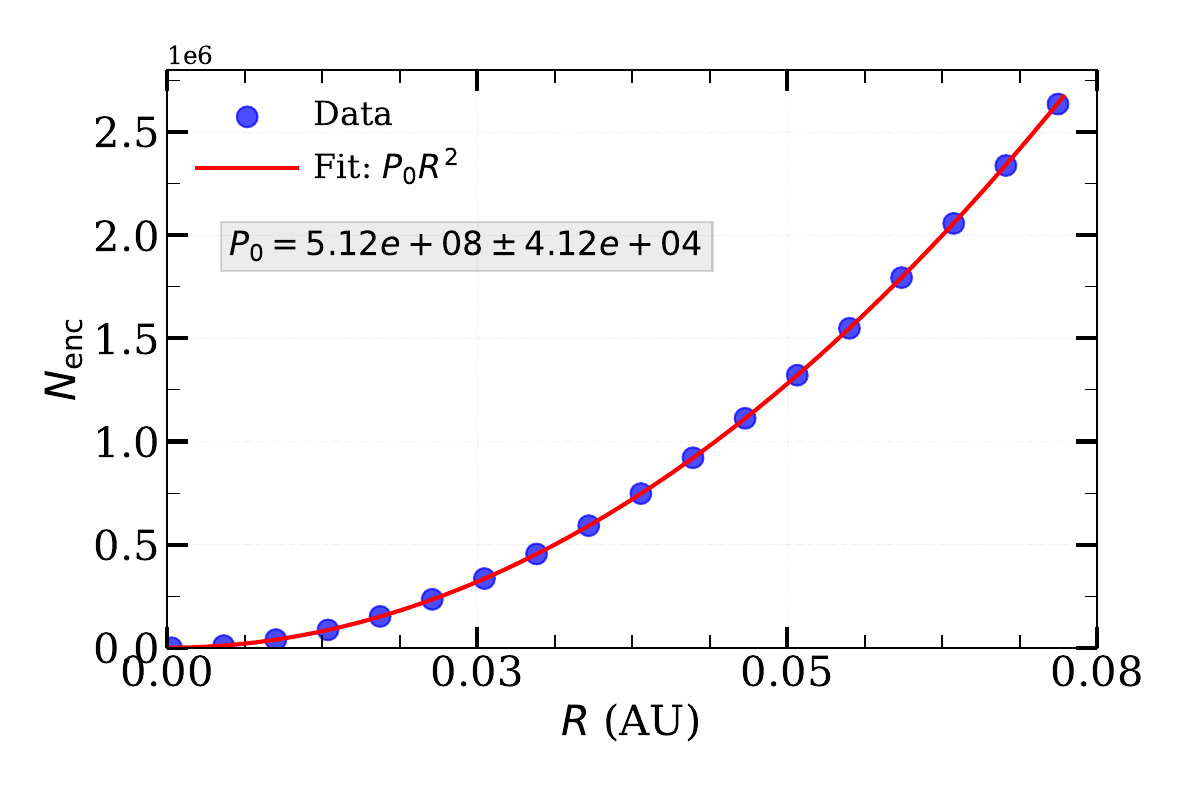}
    \caption{Cumulative number of close encounters $N(<R)$ with Mars as a function of minimum approach distance $R$ for $150~\mu$m dust grains. Blue points represent binned data, while the red curve shows the best-fit relation $N(<R) = P_0 R^2$ obtained through weighted least-squares fitting. The quadratic scaling confirms the geometric nature of encounters in three-dimensional space. The data correspond to encounters within ten times the Martian sphere of influence for $150~\mu$m dust grains.}
    \label{fig:150um_mars}
\end{figure*}

\begin{figure*}
    \centering
    \includegraphics[width=0.77\textwidth]{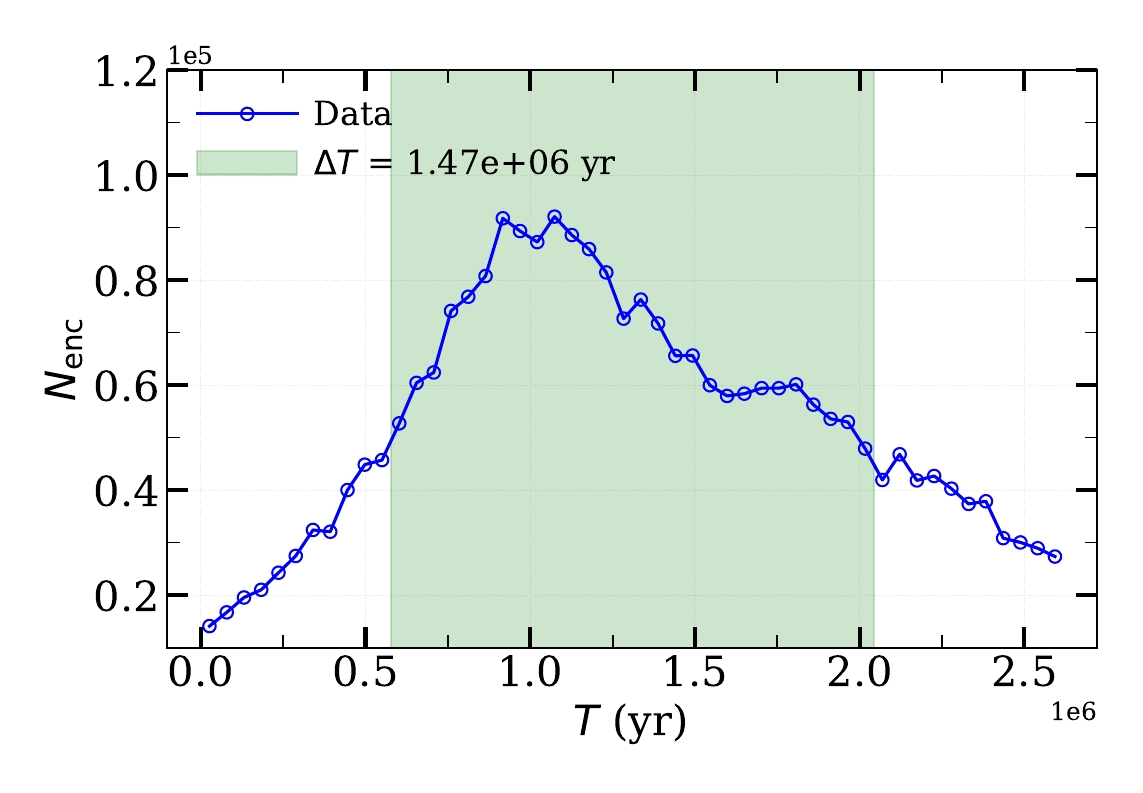}
    \caption{Temporal evolution of close encounter frequency between $150~\mu$m dust grains and Mars. Blue points denote the number of encounters per time bin, while the green shaded region highlights the effective interval ($\Delta T$), determined from the full width at half maximum (FWHM) of the distribution. This interval represents the period of statistically steady encounter rates used for flux calculation.}
    \label{fig:encounter_rate_mars}
\end{figure*}

Close encounters between dust particles and the planet were identified during the numerical integration whenever a particle entered the planet’s gravitational vicinity. Owing to the finite temporal resolution of the simulations, a single physical encounter is sampled over multiple consecutive time steps, producing a sequence of closely spaced encounter records. To avoid multiple counting of the same physical event, encounter records were grouped into individual \textit{encounter passages} based on their temporal separation. Consecutive records separated by short time intervals were treated as belonging to the same passage, whereas a new passage was defined whenever the time gap between successive records exceeded a prescribed threshold. Such large temporal gaps indicate that the particle had exited the planetary influence region before re-entering at a later epoch. For each encounter passage, the representative encounter was defined as the instant of closest approach, determined by selecting the record corresponding to the minimum particle--planet distance within that passage. The associated relative velocity at closest approach was used to characterize the encounter. This procedure ensures a one-to-one correspondence between physical encounter events and the derived encounter statistics.

The spatial distribution of close encounters reveals the fundamental geometric nature of dust-planet interactions. Figure~\ref{fig:150um_mars} presents the cumulative number of encounters $N(<R)$ with Mars as a function of the minimum approach distance $R$, for $150~\mu$m dust grains. The distribution follows the expected quadratic scaling $N(<R) = P_0 R^2$, as demonstrated by the excellent fit (reduced $\chi^2 = 1.73$, coefficient of determination $R$-squared = $1.00$). This quadratic dependence indicates that close encounters are geometrically distributed, with the cumulative encounter probability scaling with the effective cross-sectional area. The fitted parameter $P_0 = ({5.12 \pm 0.041}) \times 10^{8}$~AU$^{-2}$ encapsulates the overall normalization of the encounter distribution. The optimal binning was determined through $\chi^2$ minimization, ensuring robust statistical characterization of the spatial distribution. This $R^2$ scaling holds across the entire dynamic range from $3.5\times10^{-4}$ to $7.23\times10^{-2}$~AU, confirming the geometric nature of the encounter process at Mars. Extrapolating this relation to the physical radius of Mars, $R = R_M$, yields the fractional number of dust impacts on the planetary surface, which represents the probability that a migrating grain reaching the planet’s vicinity results in a surface impact.

The temporal distribution of close encounters exhibits a characteristic profile with distinct phases (Figure~\ref{fig:encounter_rate_mars}). After an initial rise period as when the grain begins to reach the planet, the encounter rate reaches a plateau where it remains approximately constant, followed by a gradual decline as the migrating dust population depletes. The full width at half maximum (FWHM) of this distribution defines the effective time interval $\Delta T = {1.47 \times 10^{6}}$~years during which the flux can be considered statistically steady. For Mars, $\Delta T$ represents $\approx 56\%$ of the total encounter time $T_{\text{enc}} = {2.619 \times 10^{6}}$~years, indicating sustained dust delivery over this period. The temporal FWHM method for determining $\Delta T$ offers a robust, data-driven approach for identifying the steady-state phase, avoiding arbitrary time interval selection. In our simulations particles are released at a single epoch rather than through continuous dust production. A large ensemble statistically represents the dust population, and the flux is estimated during the plateau phase. The duration of this plateau depends on particle size: smaller grains migrate faster due to stronger PR drag and thus show shorter intervals, while larger grains remain longer in the terrestrial planet region. For clarity, only a representative temporal profile is shown in Figure~\ref{fig:encounter_rate_mars}. These spatial-temporal distributions reveal a size-dependent residence time, directly influencing each grain size's contribution to the total impact flux -- a quantity that, to our knowledge, is quantified here for the first time. We also identify planet-specific spatial asymmetries in encounter patterns, reflecting differences in gravitational focusing and orbital geometry. The encounter probability per unit area, $P_0$ (tabulated for all particle sizes within the planetary encounter region in Table~\ref{tab:planet_flux}), serves as a normalization factor for comparison with theoretical predictions or other dust source populations.

\begin{figure*}[ht]
    \centering
    \includegraphics[width=0.8\textwidth]{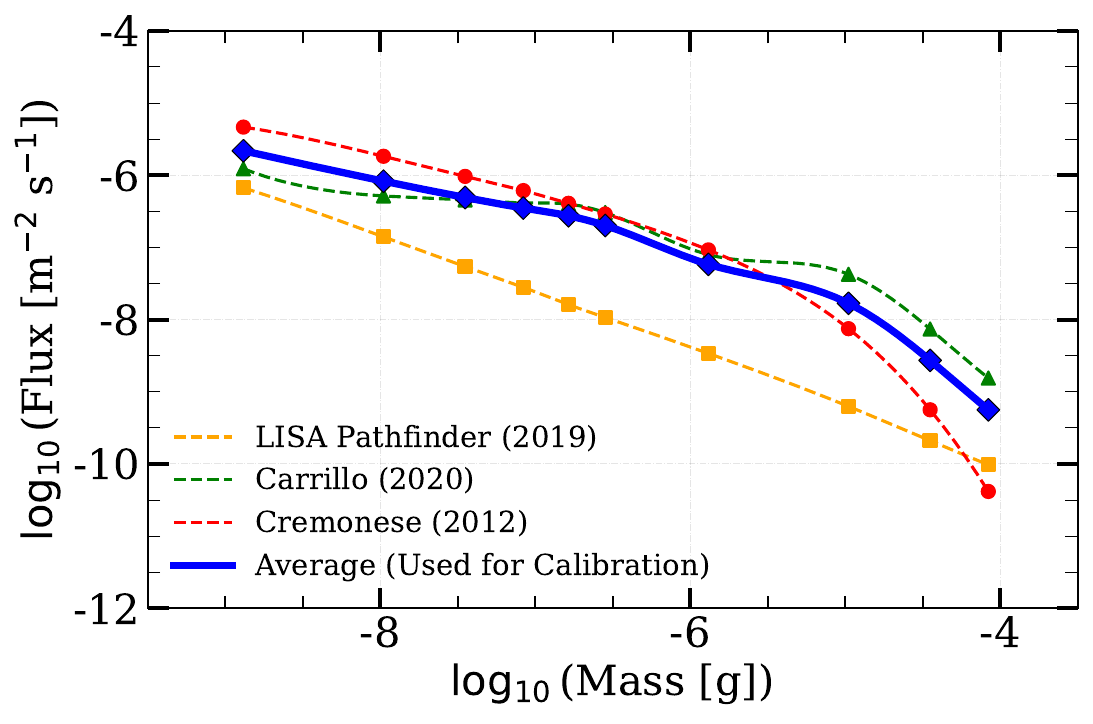}
    \caption{Size frequency distribution of Earth-based calibration data. Dashed lines show individual flux measurements from LISA Pathfinder, asteroidal flux estimate of Carrillo-Sánchez et al. (2020), and Cremonese et al. (2012). The thick solid blue line represents the average flux used as the calibration reference.}
    \label{fig:calibration_sfd}
\end{figure*}

The impact flux $g_M$ is obtained by combining the spatial and temporal statistics of close encounters through Eq.~(13). The spatial analysis yields the  fractional number of impacts $n_M$ by extending $P(R)$ to the planetary radius. This quantity is then rescaled to an effective value $n_M'$ by accounting for the fact that impacts occur only during the steady-state interval $\Delta T$ identified from the temporal distribution. For $150~\mu$m grains impacting Mars, this procedure results in a calibrated flux of $3.05 \times 10^{-10}$~m$^{-2}$~s$^{-1}$  (see Section~\ref{sec:flux_calculation}). The passage-grouping algorithm is essential for ensuring statistical robustness, as it consolidates raw encounter records into distinct physical approaches; The $50$-day gap threshold effectively distinguishes separate encounter events while ensuring that continuous approaches are treated as single passages. 
Flux calibration using a single Earth-based dataset can introduce systematic biases that propagate to all planetary flux predictions. For example, \citet{Borin2017} calibrate exclusively against Cremonese et al. (2012), leaving their results sensitive to the uncertainties of a single observational source. To mitigate this, we calibrate against three independent Earth-based datasets simultaneously: LISA Pathfinder \citep{Thorpe_2019}, asteroidal flux estimate of \citet{CARRILLOSANCHEZ2020113395}, and \citet{2012ApJ...749L..40C}. For LISA Pathfinder, we apply a gravitational focusing correction to convert the L1 flux to Earth's vicinity. The Carrillo dataset is converted from global influx to flux per unit area. For each mass bin, we compute the average of the three fluxes to obtain our Earth calibration curve, which is then used to scale simulated fluxes to Mercury, Venus, and Mars. Figure~\ref{fig:calibration_sfd} presents the size frequency distribution of the three datasets. The individual fluxes are shown as dashed lines, while the thick solid line represents the average flux used as our calibration reference. This distribution shows reasonable mutual consistency across the mass range of interest, with deviations within the observational uncertainties typical of independent measurement techniques. This multi-source approach reduces dependence on any single measurement and yields flux predictions in better agreement with independent in-situ observations/models compared to single-source calibrations.

\begin{table*}[h]
\centering
\caption{Calibrated dust flux parameters and velocities derived from close-encounter statistics for Mars, Venus, and Mercury. Listed are the particle radius, mass, fitted encounter probability coefficient $P_0$, resulting surface impact flux, mean relative encounter velocity $v_{\rm rel}$, and estimated impact velocity $v_{\rm Imp}$.}
\label{tab:planet_flux}
\vspace{3.0em}

\begin{tabular}{cccccc}
\multicolumn{6}{c}{\textbf{Mars: Calibrated dust flux and impact velocities}} \\
\toprule
Particle Radius ($\mu m$) & Mass (g) & $P_0$ (enc/AU$^2$) & Flux (m$^{-2}$ s$^{-1}$) & $v_{\text{rel}}$ (km/s) & $v_{\text{Imp}}$ (km/s) \\
\midrule
5    & $1.31 \times 10^{-9}$  & $(2.290 \pm 0.009) \times 10^{7}$  & $1.36 \times 10^{-6}$  & 5.61 & 7.69 \\
10   & $1.05 \times 10^{-8}$  & $(4.626 \pm 0.004) \times 10^{7}$  & $5.31 \times 10^{-7}$  & 5.84 & 7.88 \\
15   & $3.53 \times 10^{-8}$  & $(6.959 \pm 0.017) \times 10^{7}$  & $3.44 \times 10^{-7}$  & 5.99 & 7.99 \\
20   & $8.38 \times 10^{-8}$  & $(9.029 \pm 0.011) \times 10^{7}$  & $2.42 \times 10^{-7}$  & 6.14 & 8.10 \\
25   & $1.64 \times 10^{-7}$  & $(1.117 \pm 0.006) \times 10^{8}$  & $2.11 \times 10^{-7}$  & 6.32 & 8.25 \\
30   & $2.83 \times 10^{-7}$  & $(1.314 \pm 0.017) \times 10^{8}$  & $1.30 \times 10^{-7}$  & 6.35 & 8.28 \\
50   & $1.31 \times 10^{-6}$  & $(2.098 \pm 0.020) \times 10^{8}$  & $3.52 \times 10^{-8}$  & 6.62 & 8.47 \\
100  & $1.05 \times 10^{-5}$  & $(3.817 \pm 0.011) \times 10^{8}$  & $8.87 \times 10^{-9}$  & 7.08 & 8.84 \\
150  & $3.53 \times 10^{-5}$  & $(5.118 \pm 0.041) \times 10^{8}$  & $1.48 \times 10^{-9}$ & 7.70 & 9.34 \\
200  & $8.38 \times 10^{-5}$  & $(5.592 \pm 0.028) \times 10^{8}$  & $3.39 \times 10^{-10}$ & 8.05 & 9.64 \\
\bottomrule
\end{tabular}

\vspace{1em}

\begin{tabular}{cccccc}
\multicolumn{6}{c}{\textbf{Venus: Calibrated dust flux and impact velocities}} \\
\toprule
Particle Radius ($\mu m$) & Mass (g) & $P_0$ (enc/AU$^2$) & Flux (m$^{-2}$ s$^{-1}$) & $v_{\text{rel}}$ (km/s) & $v_{\text{Imp}}$ (km/s) \\
\midrule
5    & $1.31 \times 10^{-9}$  & $(5.202 \pm 0.002) \times 10^{7}$  & $3.55 \times 10^{-6}$  & 6.89 & 12.71 \\
10   & $1.05 \times 10^{-8}$  & $(1.076 \pm 0.003) \times 10^{8}$  & $1.39 \times 10^{-6}$  & 7.10 & 12.80 \\
15   & $3.53 \times 10^{-8}$  & $(1.638 \pm 0.004) \times 10^{8}$  & $8.54 \times 10^{-7}$  & 7.26 & 12.88 \\
20   & $8.38 \times 10^{-8}$  & $(2.180 \pm 0.005) \times 10^{8}$  & $5.52 \times 10^{-7}$  & 7.27 & 12.89 \\
25   & $1.64 \times 10^{-7}$  & $(2.749 \pm 0.005) \times 10^{8}$  & $4.38 \times 10^{-7}$  & 7.46 & 13.00 \\
30   & $2.83 \times 10^{-7}$  & $(3.367 \pm 0.006) \times 10^{8}$  & $3.19 \times 10^{-7}$  & 7.35 & 12.94 \\
50   & $1.31 \times 10^{-6}$  & $(5.610 \pm 0.007) \times 10^{8}$  & $9.21 \times 10^{-8}$  & 7.61 & 13.11 \\
100  & $1.05 \times 10^{-5}$  & $(1.034 \pm 0.047) \times 10^{9}$  & $2.43 \times 10^{-8}$  & 8.46 & 13.67 \\
150  & $3.53 \times 10^{-5}$  & $(1.254 \pm 0.074) \times 10^{9}$  & $3.64 \times 10^{-9}$ & 9.29 & 14.22 \\
200  & $8.38 \times 10^{-5}$  & $(1.182 \pm 0.076) \times 10^{9}$  & $7.25 \times 10^{-10}$ & 10.30 & 14.93 \\
\bottomrule
\end{tabular}

\vspace{1em}

\begin{tabular}{cccccc}
\multicolumn{6}{c}{\textbf{Mercury: Calibrated dust flux and impact velocities}} \\
\toprule
Particle Radius ($\mu m$) & Mass (g) & $P_0$ (enc/AU$^2$) & Flux (m$^{-2}$ s$^{-1}$) & $v_{\text{rel}}$ (km/s) & $v_{\text{Imp}}$ (km/s) \\
\midrule
5    & $1.31 \times 10^{-9}$  & $(1.793 \pm 0.493) \times 10^{8}$  & $1.28 \times 10^{-5}$  & 12.17 & 12.97 \\
10   & $1.05 \times 10^{-8}$  & $(3.398 \pm 0.680) \times 10^{8}$  & $4.55 \times 10^{-6}$  & 12.00 & 12.83 \\
15   & $3.53 \times 10^{-8}$  & $(4.741 \pm 0.736) \times 10^{8}$  & $2.51 \times 10^{-6}$  & 11.63 & 12.48 \\
20   & $8.38 \times 10^{-8}$  & $(6.206 \pm 0.808) \times 10^{8}$  & $1.78 \times 10^{-6}$  & 11.02 & 11.94 \\
25   & $1.64 \times 10^{-7}$  & $(8.038 \pm 0.917) \times 10^{8}$  & $1.46 \times 10^{-6}$  & 11.08 & 11.99 \\
30   & $2.83 \times 10^{-7}$  & $(9.876 \pm 0.984) \times 10^{8}$  & $1.00 \times 10^{-6}$  & 11.11 & 12.02 \\
50   & $1.31 \times 10^{-6}$  & $(2.008 \pm 1.454) \times 10^{9}$  & $3.64 \times 10^{-7}$  & 11.35 & 12.22 \\
100  & $1.05 \times 10^{-5}$  & $(3.438 \pm 2.072) \times 10^{9}$  & $8.40 \times 10^{-8}$  & 12.65 & 13.45 \\
150  & $3.53 \times 10^{-5}$  & $(3.447 \pm 2.173) \times 10^{9}$  & $1.15 \times 10^{-8}$  & 13.44 & 14.20 \\
200  & $8.38 \times 10^{-5}$  & $(2.638 \pm 1.901) \times 10^{9}$  & $1.63 \times 10^{-9}$ & 14.33 & 15.06 \\
\bottomrule
\end{tabular}

\end{table*}

The flux estimation illustrated for the representative $150~\mu$m grain size on Mars was applied consistently to all particle sizes and to all planets considered. The complete set of derived parameters and velocities is summarized in Table~\ref{tab:planet_flux}. The modeled dust fluxes obtained in this work can be compared with recent in--situ measurements of IDPs in near--Earth orbit. Using the Dust EXperiment (DEX) flown onboard the POEM platform of the PSLV--C58 (XPoSat) mission, \citet{Pabari2025DEX} reported time--resolved IDP fluxes in low--Earth orbit at an altitude of $\sim350$ km. The measured flux by DEX during the observation period lies in the range $\sim10^{-3}$ -- $10^{-2}$ m$^{-2}$ s$^{-1}$, with an average value of $6.46\times10^{-3}$ [$2.95 \times 10^{-3}$, $9.97 \times 10^{-3}$] m$^{-2}$ s$^{-1}$. As the flux of smaller particles dominate over larger particles, the DEX derived flux was shown concentrated towards smaller particle mass (i.e., near $10^{-18}$ kg). In contrast, the present model is calibrated using three Earth-based datasets discussed above, which correspond to larger particle masses and therefore yield lower calibrated flux values. The DEX measurements was not targeted for identification of the dust source, it may include contributions from multiple sources, whereas the present model estimates only the asteroidal component. Figure~\ref{fig:combined_flux_allcalib} presents the calibrated impact fluxes as a function of particle mass for Mercury, Venus and Mars with the average Earth calibration curve shown together for reference to illustrate the systematic dependence on particle size and planetary environment. For all bodies, the flux decreases monotonically with increasing particle mass, spanning several orders of magnitude across the considered range. This trend reflects the combined influence of size-dependent dynamical evolution and the decreasing abundance of large grains available for impact. At a given particle mass, the relative ordering of the fluxes follows the expected radial trend, with Mercury experiencing the highest impact rates, followed by Venus and Mars. Since the grain sizes considered in this work ($\sim$5--200~$\mu$m) are well within the regime where Lorentz forces are negligible, the approximately $1/r^{2}$ heliocentric flux behavior reported by \citet{Mannarticle} provides additional physical support for the flux trends obtained in our simulations. The separation between the planetary curves remains approximately constant in logarithmic space, indicating that the mass dependence of the flux is broadly similar across planets, while differences in absolute magnitude are governed by heliocentric distance, encounter probabilities, and gravitational cross sections. The smooth curves shown in Fig.~\ref{fig:combined_flux_allcalib} are intended only as visual aids to emphasize the global behavior of the flux with particle mass. The underlying results correspond exclusively to the discrete grain sizes explicitly simulated.

\begin{figure*}[ht]
    \centering
    \includegraphics[width=0.8\textwidth]{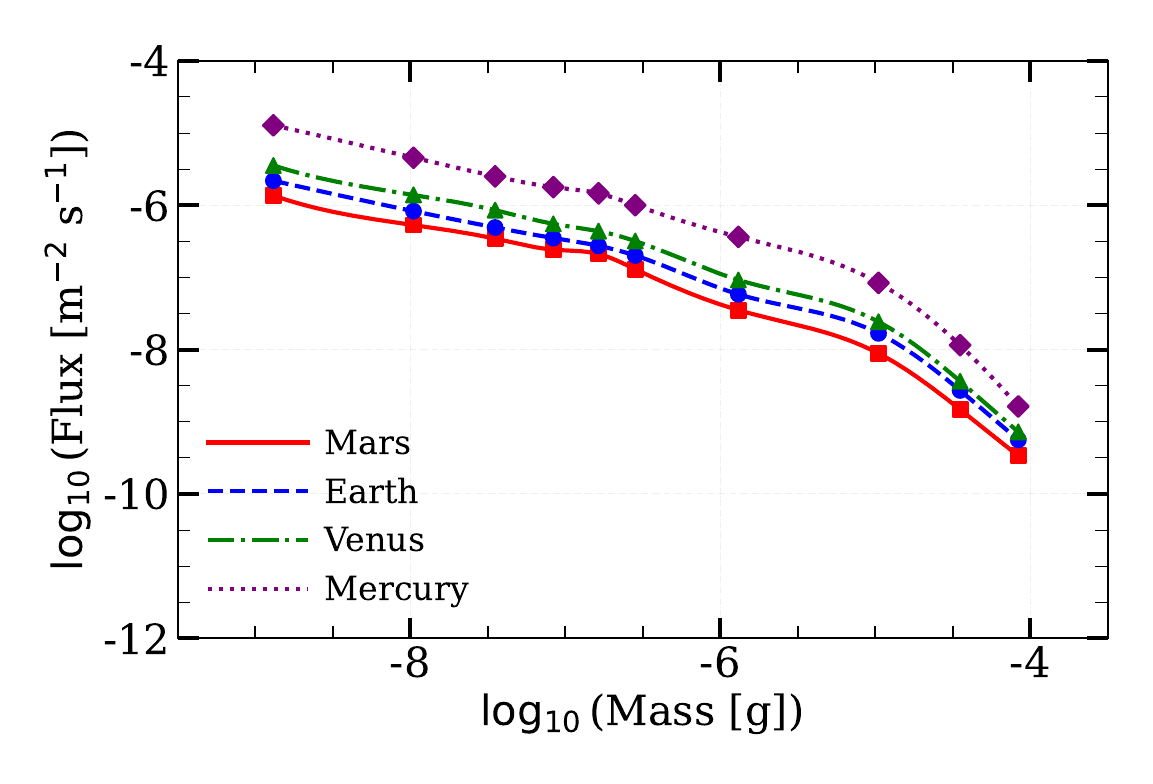}
    \caption{Simulated interplanetary dust particle flux as a function of particle mass for the terrestrial planets Mercury, Venus, Earth, and Mars. Solid and dashed curves represent spline-interpolated trends, while symbols denote discrete simulation outputs.}
    \label{fig:combined_flux_allcalib}
\end{figure*}

To assess the realism of our simulated Mars impact fluxes, we compared them with  two independent references: (1) the scaled Grün et al. (1985) interplanetary dust model at Mars with gravitational focusing, and (2) flux estimates derived from the Langmuir Probe and Waves (LPW) instrument on the MAVEN spacecraft \citep{Andersson2015Sci...350.0398A, PABARI201868}. The MAVEN-based flux estimates were derived using the model proposed by \citet{PABARI20171}, which expresses the cumulative number of interplanetary dust particles larger than radius $s$ as

\begin{equation}
N(s) = 5 \left( \frac{17}{s} \right)^{1.32},
\end{equation}
where the power-law index was empirically determined from LPW observations. In Eq.~(20), the symbol $s$ is used instead of $R$ to maintain consistent notation throughout the present work. Our simulation shows excellent agreement with the Grün model, with a mean offset of only $0.04$ orders of magnitude (a factor of $1.09$ higher) and a scatter of $0.20$ orders of magnitude. This close agreement validates our Earth-calibrated asteroidal source function. In contrast, the MAVEN-inferred fluxes are systematically higher. Our simulation is lower by approximately $1.45$ orders of magnitude compared to the MAVEN estimates, with a scatter of $0.41$ orders of magnitude. This discrepancy is expected, as the MAVEN estimates likely include contributions from additional dust sources (e.g., cometary dust, impact ejecta from Phobos and Deimos) not included in our asteroidal-only model. Previous dynamical and observational studies \citep{2010ApJ...713..816N, 2018ApJ...863...31P} have examined these populations. MAVEN flux measurements cover the mass range $10^{-11}~\mathrm{g} \lesssim m \lesssim 10^{-1}~\mathrm{g}$, while our simulations span $10^{-9}~\mathrm{g} \lesssim m \lesssim 10^{-5}~\mathrm{g}$. Extrapolating flux–mass trend derived from the simulations toward smaller particle masses (down to $m \sim 10^{-10}~\mathrm{g}$) predicts fluxes of the order of $10^{-4}~\mathrm{m^{-2}\,s^{-1}}$, which is consistent with the MAVEN LPW measurements ($\sim 10^{-4}$ m$^{-2}$ s$^{-1}$) and the model estimates reported by \citet{PABARI20171}. Extrapolating our Mars flux trend to the same mass threshold as Jorgensen's $\geq 1~\mu$m gives a cumulative flux of $2.8 \times 10^{-5}$~m$^{-2}$~s$^{-1}$, close to Jorgensen's reported $2.56 \times 10^{-5}$ at 1.6~AU. Overall, these results confirm that the simulation reliably reproduces the observed mass dependence of the asteroidal dust flux at Mars, providing confidence in the model for further investigation of the Martian dust environment.

The realism of our simulated Venus dust fluxes was assessed by comparing with multiple independent datasets and models: the scaled \citet{GRUN1985244} interplanetary dust model at Venus, the \citet{CARRILLOSANCHEZ2020113395} flux estimates,  the \citet{Borin2017} dataset, in-situ observations from Galileo and Helios spacecraft, and the polynomial model of \citet{PABARI2023105617}. Our simulation shows excellent agreement with the scaled Grün model at Venus, with a mean offset of only $0.09$ orders of magnitude and a scatter of $0.19$ orders of magnitude. Agreement with Carrillo-Sánchez et al. (2020) is also good ($0.22$ orders of magnitude offset). The \citet{Borin2017} dataset lies significantly lower, with our simulation higher by $1.10$ orders of magnitude. Notably, the in-situ Galileo and Helios observations are $0.58$ and $0.65$ orders of magnitude above the Grün model, respectively, placing them closer to our simulation than to the Borin dataset. This indicates that our simulation better reproduces the actual spacecraft measurements. The superior performance of our model compared to \citet{Borin2017} likely stems from our multi-dataset calibration approach. Unlike \citet{Borin2017}, who calibrated using a single Earth-based  \citet{2012ApJ...749L..40C} flux curve, the present work employs a multi-source calibration approach, reducing systematic biases and providing a more robust normalization that results in flux predictions aligning more closely with in-situ observations at Venus. Our simulation also follows the same mass dependence as the Pabari polynomial model for Venus, which was constructed to fit multiple Venus observations and models, though with a systematic offset of $0.55$ orders of magnitude. Given the excellent agreement with the scaled Grün model at Venus — the widely accepted standard for IDP flux — and the good consistency with Galileo and Helios observations, we conclude that our calibrated asteroidal model reliably represents the Venus dust environment. Nevertheless, our multi-dataset calibration represents a significant improvement over previous single-dataset approaches and provides a more observationally grounded estimate of the Venus dust flux.

We compare our simulated dust flux near Mercury with the model of \citet{MULLER20021101}, which is based on Divine's five population model that explicitly includes an asteroidal population and was empirically constrained using multiple dust flux measurements (Grün et al. 1985; Pioneer 10/11; Helios; Ulysses; Galileo). For particle masses in the range $\sim 10^{-9}$--$10^{-4}$ g, the Müller model predicts fluxes between $\sim 4\times10^{-5}$ and $\sim 10^{-9}$ m$^{-2}$ s$^{-1}$. Our calibrated simulation yields fluxes between $\sim 1.28\times10^{-5}$ and $\sim 1.63\times10^{-9}$ m$^{-2}$ s$^{-1}$ over the same mass range (see Table~\ref{tab:planet_flux}). The comparison shows excellent agreement between our simulation and the Müller model. The median offset is only $-0.04$ orders of magnitude (a factor of $0.91$, i.e., $9\%$ lower), with a scatter of $0.27$ orders of magnitude. This close agreement validates our Earth-calibrated asteroidal source function for Mercury, demonstrating that our calibration approach successfully extends to the innermost planet and provides a reliable estimate of the dust flux at Mercury.

Over the full simulated mass range ($10^{-9}$–$10^{-4}$~g), the surface impact flux on each planet is well described by a single power-law relation,
\begin{equation}
F(m) = C\, m^{\alpha},
\end{equation}
where $F$ is the impact flux in m$^{-2}$~s$^{-1}$, $m$ is the particle mass in
grams, $\alpha$ is the fitted slope in log--log space, and $C$ is a
planet-dependent normalization constant. A least-squares fit performed in
log--log space over the full mass interval yields
\begin{equation}
\left\{
\begin{aligned}
F_{\rm Mars}(m)    &= 10^{-11.95}\, m^{-0.73}, \quad (R^2 = 0.938),\\
F_{\rm Earth}(m)   &= 10^{-11.63}\, m^{-0.71}, \quad (R^2 = 0.939),\\
F_{\rm Venus}(m)   &= 10^{-11.59}\, m^{-0.74}, \quad (R^2 = 0.937),\\
F_{\rm Mercury}(m) &= 10^{-11.20}\, m^{-0.75}, \quad (R^2 = 0.925).
\end{aligned}
\right.
\label{eq:flux_powerlaw_planets}
\end{equation}
for Mars, Earth, Venus and Mercury respectively, with R-squared values given in parentheses. The close similarity of the fitted slopes for all four planets indicates that the mass dependence of the impact flux is largely controlled by the intrinsic dust size distribution and transportp.rocesses, rather than by planet-specific properties. The high $R^2$ values ($>0.92$) demonstrate that the power-law model provides an excellent fit to the simulated fluxes across the entire mass range for each planet. While flux estimates for individual planets exist in the literature, power-law fits calibrated against multiple independent Earth observations have not been previously provided simultaneously for all four inner planets. These fits therefore offer the community a simple, observationally-grounded tool for flux estimation across the inner Solar System. As discussed in Section~\ref{sec:style}, P-R drag dominates over collisions for our grain sizes ($10^{-9}$–$10^{-4}$ g); therefore, mutual collisions do not significantly affect our results.}
 
\subsection{Velocity Distribution at different planets}  

\begin{figure*}[ht]
    \centering
    \includegraphics[width=0.8\textwidth]{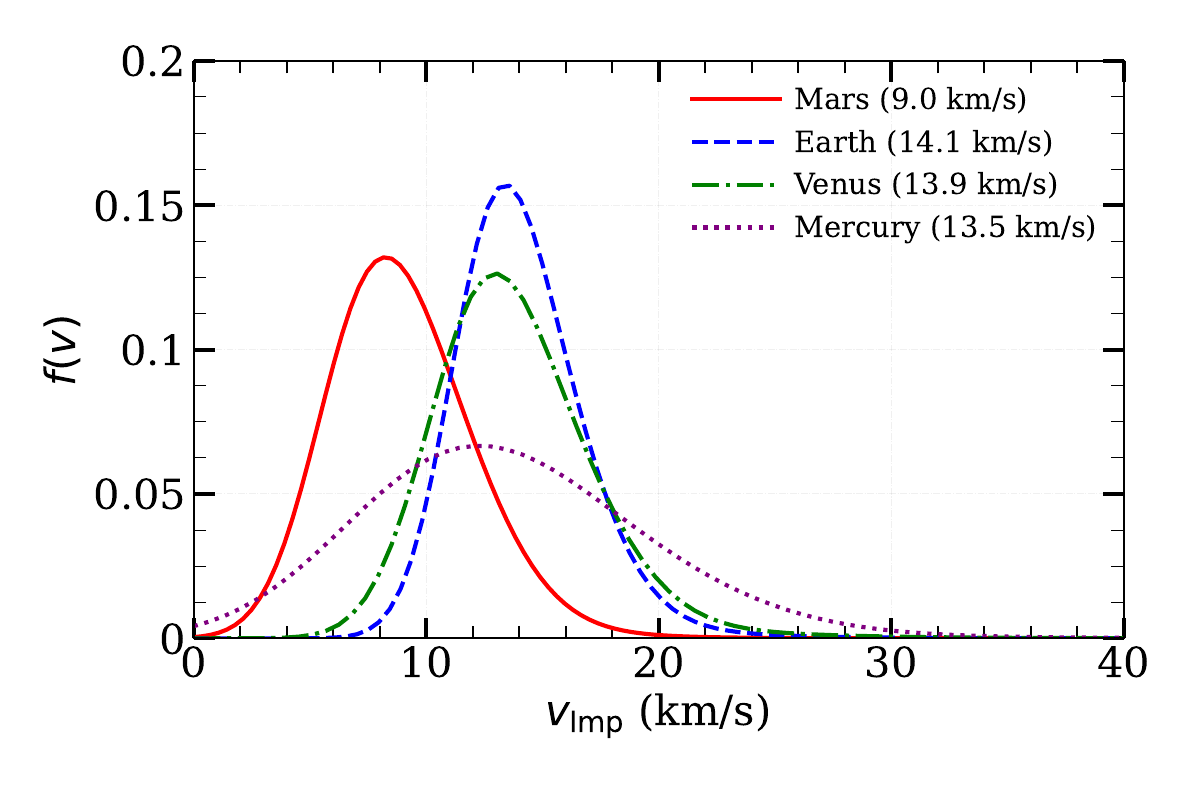}
  \caption{Normalized impact velocity distributions of dust particles impacting Mercury, Venus, Earth, and Mars, obtained from N-body simulations. The curves represent kernel density estimates of the relative impact velocities at each planet.}
    \label{fig:velocity_dist}
\end{figure*}

Our dynamical simulations reveal systematic variations in impact velocities across the terrestrial planets (Fig.~\ref{fig:velocity_dist}). The mean impact velocities follow a clear hierarchy: Earth (14.1 km/s) $>$ Venus (13.9 km/s) $>$ Mercury (13.5 km/s) $>$ Mars (9.0 km/s). For Mars, dust particles originating from the main asteroid belt experience relatively weak dynamical perturbations, and their velocity distribution remains close to the original source population, resulting in a low characteristic impact velocity of $\sim 9$~km~s$^{-1}$. In contrast, particles impacting the Earth exhibit a significantly higher impact speed, with a mean value of $\sim 14.1$~km~s$^{-1}$. For reference, LDEF-based experimental analyses of near-Earth meteoroids typically adopt a normal impact velocity component of order $v_n \sim 12$~km~s$^{-1}$ for Earth, assuming an average incidence angle of $45^\circ$ \citep{doi:10.1126/science.262.5133.550}. Our elevated velocity is a direct consequence of the dynamical excitation of particle eccentricities during inward migration. The Earth impact velocity shows a clear size dependence (see Table~\ref{tab:planet_flux}): larger grains ($\gtrsim 100~\mu$m) impact at higher speeds, while smaller grains impact more slowly. This trend arises because larger grains undergo prolonged MMR trapping with Earth (Fig.~\ref{fig:dynamics_doublepanel}), during which their eccentricities are periodically pumped, maintaining high orbital velocities. Smaller grains migrate faster via PR drag, experience only brief resonance captures, and therefore exhibit lower eccentricities and impact speeds. At Venus, the characteristic impact velocity decreases slightly to $\sim 13.9$~km~s$^{-1}$, which may be attributed to progressive orbital circularization driven by PR drag. Earth ($\sim 14.10$~km~s$^{-1}$) and Venus ($\sim 13.9$~km~s$^{-1}$) show the highest impact velocities, resulting from the combination of their substantial gravitational focusing (escape velocities of $11.19$~km~s$^{-1}$ and $10.36$~km~s$^{-1}$, respectively) with moderate relative velocities at 1~AU. A further modest reduction is observed at Mercury, where the mean impact velocity is $\sim 13.5$ km s$^{-1}$. This value averages simulated impact velocities over all orbital phases, naturally accounting for Mercury's moderate orbital eccentricity. Comparable impact speeds have been reported in previous studies e.g., \citet{Cintala1992JGR....97..947C}; \citet{CARRILLOSANCHEZ2020113395}, although the latter find somewhat lower velocities for asteroidal particles of about 12.0 km s$^{-1}$ at Earth, 6.5 km s$^{-1}$ at Mars, and 11.4 km s$^{-1}$ at Venus.

\label{sec:velocities_results}

\begin{figure*}[ht]
    \centering
    \includegraphics[width=0.9\textwidth]{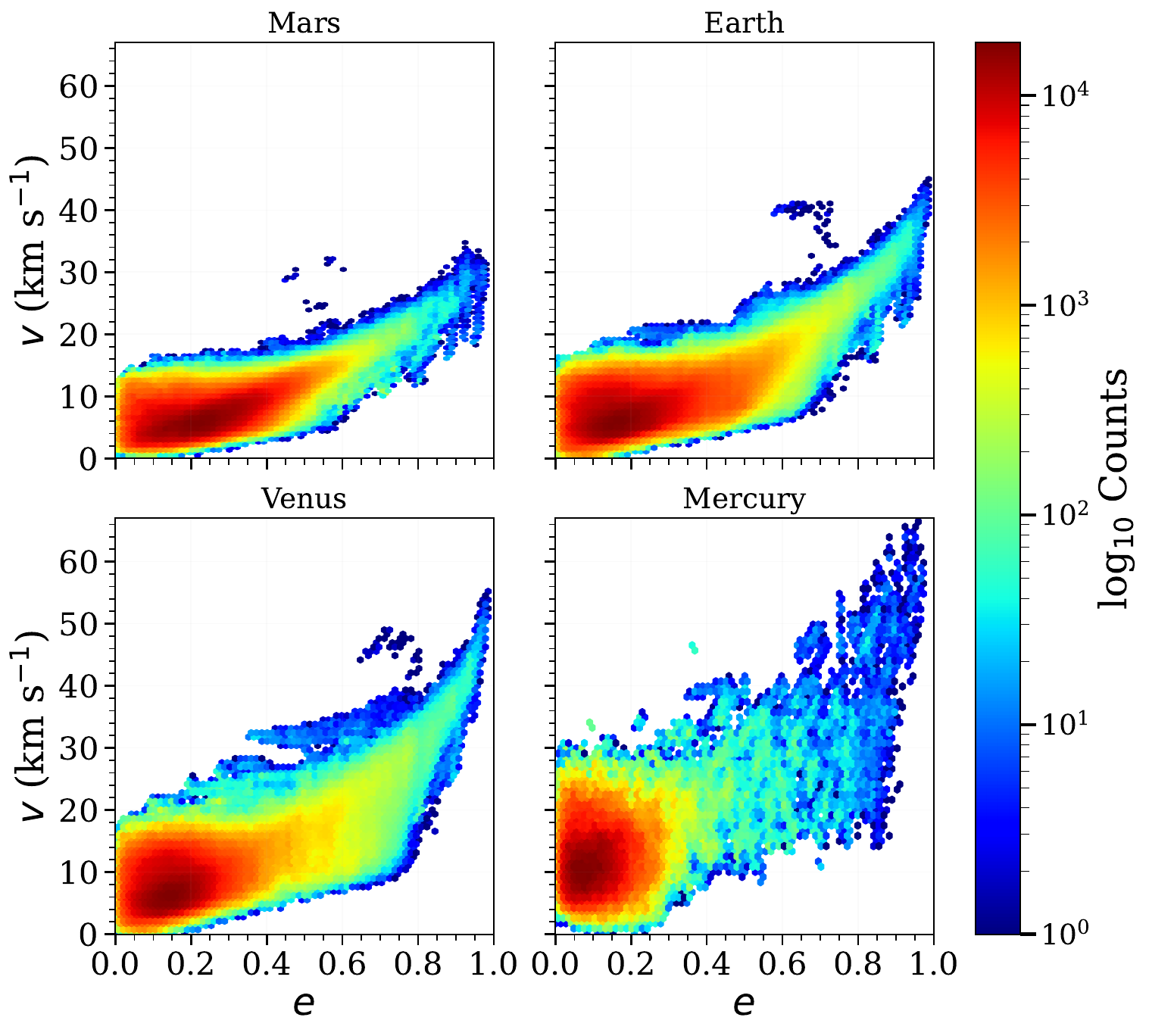}
  \caption{Hexbin maps of velocity $v$ as a function of orbital eccentricity $e$ for IDPs impacting (a) Mars, (b) Earth, (c) Venus, and (d) Mercury. Colors indicate the logarithmic number of impact events per bin. A systematic increase in both the characteristic velocity and the velocity dispersion is observed toward the inner Solar System, together with the emergence of a dynamically excited high-eccentricity population that contributes disproportionately to the high-velocity tail.
}
    \label{fig:e_vimp_ecc}
\end{figure*}

\begin{figure*}[ht]
    \centering
    \includegraphics[width=0.9\textwidth]{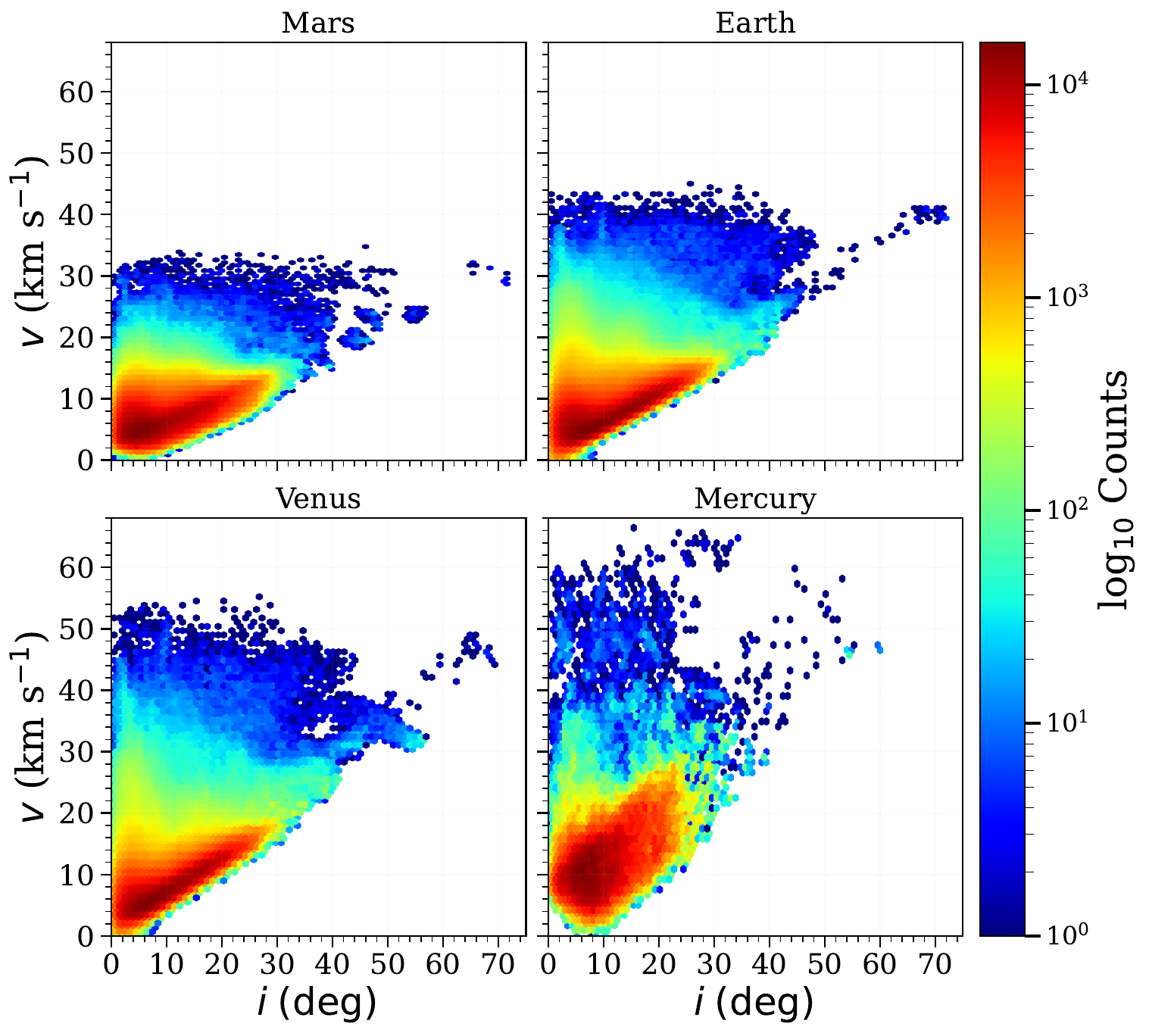}
  \caption{Hexbin maps of velocity $v$ as a function of orbital inclination $i$ for IDPs impacting (a) Mars, (b) Earth, (c) Venus, and (d) Mercury. Colors indicate the logarithmic number of impact events per bin.}
    \label{fig:i_vimp_inc}
\end{figure*}

\begin{figure*}[ht]
    \centering
    \includegraphics[width=0.9\textwidth]{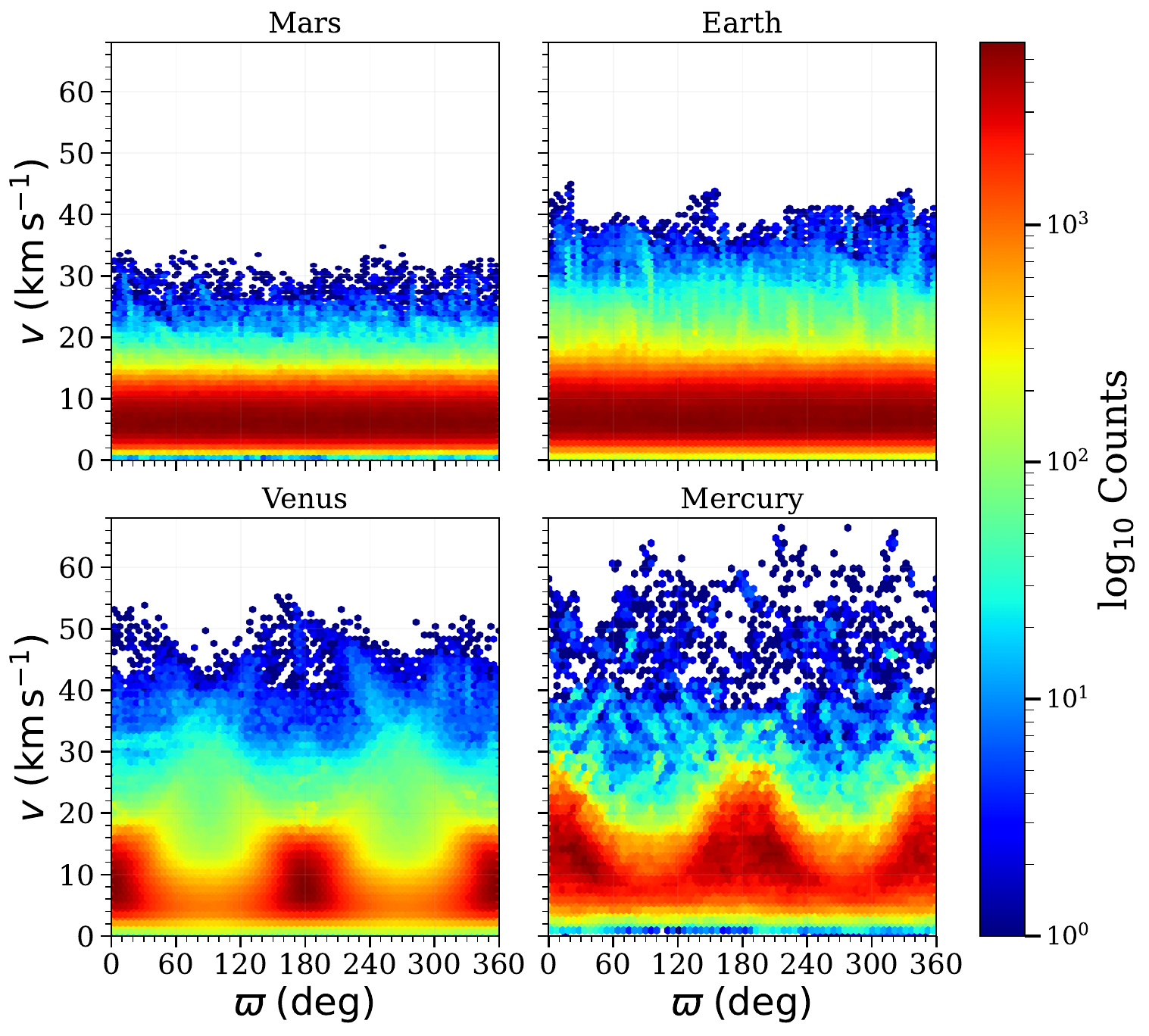}
    \caption{Distribution of dust–planet encounter velocities as a function of the longitude of perihelion, $\varpi = \Omega + \omega$, for Mercury, Venus, Earth, and Mars. Colors indicate the logarithmic number of close encounters in each hexagonal bin. While the distributions for Earth and Mars are nearly uniform in $\varpi$, pronounced structure is evident for Venus and especially Mercury, where high-velocity encounters preferentially occur for specific apsidal orientations.}
    \label{fig:w_vimp_apsidal}
\end{figure*}

Table~\ref{tab:planet_flux} lists the computed impact velocities ($v_{\rm imp}$) and relative velocities ($v_{\rm rel}$) for all 10 dust particle sizes considered in our simulations. A valuable point of comparison is provided by \citet{1992Icar...97...70J}, who reported that asteroidal dust grains with radii of 10, 30, and 100~$\mu$m have mean eccentricities of $e \simeq 0.10$, $0.11$, and $0.13$, respectively, and cross the Earth’s orbit with relative velocities of approximately 5, 5, and 6~km~s$^{-1}$. In our simulations, the corresponding mean relative velocities at Earth are slightly higher, with values of $\sim 6.41 $~km~s$^{-1}$ for 10~$\mu$m grains, $\sim 6.95 $~km~s$^{-1}$ for 30~$\mu$m grains, and $\sim 7.79$~km~s$^{-1}$ for 100~$\mu$m grains. Although impact velocity increases modestly with particle mass, the variation across all 10 sizes for a single planet is relatively small. We therefore combine the results for all particle sizes to compute a single representative impact velocity for each planet (Table~\ref{tab:velocity_stats}), following \citet{2009A&A...503..259B}, who showed that velocity distributions are broadly similar across particle sizes. We note, however, that both our model and that of \citet{2009A&A...503..259B} neglect interparticle collisions, which recent studies suggest can modify orbital element distributions and thus impact velocities \citep{Pokorny2018article, 2024PSJ.....5...82P}. The comparison between these size-dependent trajectories further highlights how resonance interactions and eccentricity evolution critically modulate the characteristic impact velocities of dust particles at inner planets, reported in Table ~\ref{tab:velocity_stats}. All distributions exhibit positive skewness, with means exceeding medians by 0.4--0.7 km/s (Table~\ref{tab:velocity_stats}). Mercury displays the broadest velocity range (4.3 -- 66.6 km/s), while Mars shows the narrowest (5.0 -- 35.1 km/s). Notably, the minimum impact velocities for Mars (5.0~km/s) and Mercury (4.3~km/s) closely match their respective escape velocities (5.0~km/s and 4.2~km/s), confirming that impactors must exceed escape velocity to reach the surface. In contrast, Earth and Venus have minimum impact velocities (11.2 km/s and 10.4 km/s) substantially higher than their escape velocities (11.2 km/s and 10.4 km/s), indicating additional velocity components from orbital dynamics.  These results demonstrate that impactor velocities are not simply proportional to planetary escape velocity, but emerge from the complex interplay between the dust particles' orbital distribution, the target planet's heliocentric position, and gravitational acceleration during final approach. 

\begin{deluxetable*}{lccccccc}
\tabletypesize{\footnotesize}
\setlength{\tabcolsep}{3pt}
\tablecaption{Velocity characteristics of asteroidal dust at terrestrial planets\label{tab:velocity_stats}}

\tablehead{
\colhead{Planet} &
\colhead{$\langle v_{\rm rel}\rangle$} &
\colhead{$\langle v_{\rm imp}\rangle$} &
\colhead{$\tilde{v}_{\rm imp}$} &
\colhead{$v_{\rm esc}$} &
\multicolumn{3}{c}{Fit parameters} \\
&
\colhead{(km s$^{-1}$)} &
\colhead{(km s$^{-1}$)} &
\colhead{(km s$^{-1}$)} &
\colhead{(km s$^{-1}$)} &
\colhead{$k$} &
\colhead{$\alpha$} &
\colhead{$R^2$}
}

\startdata
Mercury & $12.64 \pm 5.20$ & 13.5 & 12.8 & 4.25 &
$1.63\times10^{2} \pm 2.89\times10^{1}$ & $0.544 \pm 0.013$ & 0.98 \\
Venus   & $8.73 \pm 4.03$  & 13.9 & 13.2 & 10.36 &
$1.35\times10^{1} \pm 4.26$ & $0.349 \pm 0.024$ & 0.93 \\
Earth   & $8.14 \pm 3.50$  & 14.1 & 13.6 & 11.19 &
$3.74 \pm 1.30$ & $0.239 \pm 0.026$ & 0.88 \\
Mars    & $7.24 \pm 2.99$  & 9.0  & 8.6  & 5.03  &
$3.55\times10^{1} \pm 9.75$ & $0.299 \pm 0.020$ & 0.91 \\
\enddata

\tablecomments{
$\langle v_{\rm rel}\rangle$ is the mean relative speed prior to gravitational focusing.
$\langle v_{\rm imp}\rangle$ and $\tilde{v}_{\rm imp}$ are the mean and median impact speeds at the planetary surface, respectively,
and $v_{\rm esc}$ is the planetary escape speed.
The parameters $k$ and $\alpha$ are obtained by fitting the analytical distribution of \citet{Cintala1992JGR....97..947C};
$R^2$ denotes the coefficient of determination.
}
\end{deluxetable*}

To assess whether the simulated impact velocity distributions are consistent with analytical expectations, we fit the numerical results using the functional form originally proposed by \citet{Cintala1992JGR....97..947C}. Rather than deriving a new velocity distribution, we adopt the Cintala formulation and treat the normalization constant $k$ and the exponential decay parameter $\alpha$ as free parameters, allowing them to vary for each planet. This approach enables a direct and quantitative comparison between the numerical simulations and the analytical model. Specifically, the velocity distribution is written as
\begin{equation}
\begin{aligned}
f(v) =\;& k\, r^{0.2}
\left[
\frac{v}{\sqrt{r\left(v^{2}-v_{\mathrm{Pe}}^{2}\right)+v_{Ee}^{2}}}
\right]^3 \\
&\times
\exp\!\left[
-\alpha \sqrt{r\left(v^{2}-v_{\mathrm{Pe}}^{2}\right)+v_{Ee}^{2}}
\right],
\end{aligned}
\end{equation}
where $r$ is the heliocentric distance of the planet in AU, $v$ is the impact velocity, $v_{\mathrm{Pe}}$ is the planetary escape velocity, and $v_{Ee}=11.1~\mathrm{km~s^{-1}}$ is the Earth escape velocity at 100 km altitude. The parameters $k$ and $\alpha$ are determined by fitting this expression to the normalized velocity distributions obtained from the simulations; the resulting best-fit values, along with the corresponding coefficients of determination ($R^2$), are listed in Table~\ref{tab:velocity_stats}. The fitted parameters for Mars, Venus, Earth, and Mercury yield coefficients of determination $R^2$ in the range 0.88–0.98, indicating that the analytical form proposed by Cintala (1992) provides a good representation of the simulated impact velocity distributions when the parameters are adjusted appropriately. We note, however, that both the present model and the work of Cintala (1992) neglect interparticle collisions within the zodiacal cloud. More recent studies suggest that collisions can modify velocity distributions at terrestrial planets \citep{Pokorny2018article, 2024PSJ.....5...82P}.

We present the first joint characterization of dust–planet encounter velocities as functions of orbital eccentricity, inclination, and longitude of perihelion with impact flux simultaneously encoded as the colorbar -- for all four inner planets within a single self-consistent N-body framework. This plot architecture directly and visually demonstrates a flux–velocity decoupling: the flux-dominant population (low-$e$, low-$i$) and the velocity-dominant population (high-$e$, high-$i$) occupy distinct regions in orbital element space. Such joint characterization of velocity and flux in orbital element space is inaccessible to prior analytical formulations (e.g., Cintala 1992) and has not been presented in previous numerical studies.

Figures~\ref{fig:e_vimp_ecc} and~\ref{fig:i_vimp_inc} present a comparative and statistically robust characterization of dust–planet encounter velocities as functions of orbital eccentricity and inclination, respectively, for Mars, Earth, Venus, and Mercury. The large statistics obtained from N-body integrations reveal distinct, planet-dependent regimes that cannot be captured by analytical estimates. The eccentricity–velocity distributions (Figure~\ref{fig:e_vimp_ecc}) demonstrate that low-eccentricity grains dominate the impact flux at all terrestrial planets, forming a dense population with modest encounter velocities, characteristic of weakly excited dust populations evolving under PR drag. In contrast, a dynamically distinct high-eccentricity population produces a pronounced high-velocity tail dominating the extreme end of the encounter-velocity distribution. The extent of this high-velocity regime increases systematically toward the Sun, with Mercury exhibiting both the broadest velocity spread and the highest encounter speeds, while Mars shows a more compact distribution. This radial trend reflects the combined effects of increasing orbital velocities and enhanced dynamical excitation in the inner Solar System. The inclination–velocity distributions (Figure~\ref{fig:i_vimp_inc}) further demonstrate that encounter velocity is strongly coupled to the vertical dynamical state of the dust population. Low-inclination particles dominate the low-velocity regime at all planets, whereas highly inclined grains exhibit systematically higher encounter velocities. This behavior indicates that three-dimensional dynamical excitation, driven by resonant transport and gravitational scattering during inward migration, plays a key role in shaping the high-velocity impact population. The persistence of high-velocity features at moderate to large inclinations across all planets demonstrates that encounter velocities depend sensitively on the underlying orbital excitation of the dust population. These analyses reveal a clear decoupling between dust impact flux and impact velocity at the terrestrial planets. While the total flux is dominated by low-eccentricity, low-inclination grains evolving under PR drag, the extreme end of the impact-velocity distribution is controlled by a comparatively small, dynamically excited population with larger eccentricities and inclinations. This decoupling emerges directly from the simulation data, showing that a single characteristic impact velocity is insufficient to represent the full dynamical range of dust–planet interactions. As a consequence, surface processes controlled primarily by mass delivery (e.g., regolith gardening) are governed by low-e, low i population, whereas energy-driven processes such as impact vaporization, sputtering, and exosphere generation are dominated by a comparatively small fraction of high-e, high-i grains. This two-population regime demonstrates that flux and velocity play fundamentally different roles in dust–planet interactions and should be treated separately when assessing impact-driven surface and exospheric processes.

Figure~\ref{fig:w_vimp_apsidal} reveals a clear planet-dependent dependence of dust–planet encounter velocities on the longitude of perihelion, $\varpi = \Omega + \omega$  (see the schematic definition of these Keplerian elements in \citet{BATYGIN20191}). For Earth and Mars, the encounter velocity distribution is nearly uniform in $\varpi$, indicating that apsidal orientation plays only a minor role in shaping encounter energetics. In contrast, Venus and Mercury exhibit pronounced non-uniformity, with enhanced encounter rates and elevated impact velocities concentrated near specific apsidal orientations, notably near aligned and anti-aligned configurations ($\varpi \approx 0^\circ$ and $180^\circ$). This behavior reflects a transition from dynamically averaged encounters at larger heliocentric distances to geometrically selective encounters in the inner solar system. For the outer terrestrial planets, dust grains typically encounter the planet over a wide range of orbital phases after significant apsidal precession, effectively erasing any memory of the original perihelion orientation. Toward the inner solar system, however, encounters increasingly occur near the dust perihelion, where orbital velocities are highest and encounter geometry becomes strongly sensitive to the orientation of the apsidal line. As a result, apsidal alignment acts as a geometric filter that selectively enhances high-velocity encounters at Venus and, most strongly, at Mercury. 

Taken together, these analyses demonstrate that dust–planet encounter energetics are controlled by a combination of orbital excitation and heliocentric geometry, rather than by encounter flux alone. The results from all three figures establish that mean velocities are insufficient to characterize dust–planet interactions. Instead, the full velocity distribution—particularly its high-velocity tail—must be considered when assessing impact-driven surface modification and exosphere generation. In this framework, Mercury experiences a sustained bombardment by dynamically excited, high-velocity grains, while Earth- and Mars-crossing dust populations are dominated by lower-velocity encounters with only episodic contributions from highly energetic impacts. This fundamental contrast in bombardment regimes, shaped by radial trends in orbital dynamics and encounter geometry, naturally accounts for the divergent surface and exospheric environments observed across the inner Solar System.

\section{Conclusion} \label{sec:conclusion}
In this work, we investigated the dynamical evolution, impact fluxes, and encounter velocities of asteroidal dust using N-body simulations that incorporate radiation forces, solar wind, P-R drag, and planetary perturbations. Our results demonstrate a strong and systematic size dependence of dust dynamics, which governs both orbital evolution and contributions to planetary impact environments. Radiation pressure plays a dominant role for submicron grains, leading to dynamical blowout below $s \simeq 0.1~\mu$m, while micron-sized and larger particles remain gravitationally bound and populate the terrestrial planet region. Our analysis of impact velocity distributions reveals systematic variations across the terrestrial planets. The mean impact velocities follow a clear hierarchy: Earth (14.1 km/s) $>$ Venus (13.9 km/s) $>$ Mercury (13.5 km/s) $>$ Mars (9.0 km/s). For Mars, dust particles originating from the main asteroid belt experience relatively weak dynamical perturbations, resulting in a low characteristic impact velocity. In contrast, particles impacting Earth exhibit significantly higher impact speeds due to dynamical excitation of particle eccentricities during inward migration, particularly through mean-motion resonance trapping. At Venus, the velocity decreases slightly due to progressive orbital circularization driven by PR drag. Mercury's mean impact velocity, despite its weak gravity, remains comparable to Earth and Venus, indicating that heliocentric orbital dynamics dominate impact energetics at small heliocentric distances. \\

In a nutshell, the following contributions are made through this work.

(a) Multi-source calibration of our simulated fluxes has been carried out satisfying the three independent Earth-based observations: LISA Pathfinder (Thorpe et al., 2019), asteroidal flux of Carrillo-Sánchez et al. (2020), and Cremonese et al. (2012) maximally, which improves upon previous works that used only a single calibration source. This approach reduces the systematic biases inherent in single-dataset calibrations and provides a robust observational baseline. Validation against independent observations shows excellent agreement: our simulation matches the scaled Grün model for Mars within $0.04$ orders of magnitude and for Venus within $0.09$ orders, and agrees with the Müller (2002) model for Mercury within $0.04$ orders. Comparisons with in-situ Galileo and Helios Venus data also show consistency within $0.1$--$0.2$ orders of magnitude. These results confirm that our calibrated asteroidal model reliably reproduces the observed dust flux environment across the inner solar system, outperforming previous single-source calibrations.

(b) Analyses of semi-major axis evolution of individual dust particles reveal temporary slowing of inward migration at $\sim$1.3~AU, $\sim$1.4~AU, and $\sim$0.85~AU, identified as the 2:3, 3:5, and 4:3 mean-motion resonances with Earth, respectively, by computing mean-motion ratios using Kepler's third law. Unlike prior studies that characterize MMR trapping statistically through surface density maps and semi-major axis histograms \citep{Sommer2020A&A...635A..10S}, or present size-dependent semi-major axis evolution without resonance identification \citep{Borin2017}, the present work provides the first trajectory-level identification of specific MMR locations for dust population evolving under P-R drag. These trapping events temporarily excite particle eccentricities and the features near 1.3--1.4~AU are explicitly verified to be non-resonant with Mars. This provides direct dynamical evidence for resonant dust structures around Earth in the inner Solar System.

(c) For the first time, the relative velocity distributions of impactors at each inner planet as a function of orbital elements (i.e., eccentricity, inclination, and longitude of perihelion) are presented. Our results show decoupling between dust impact flux and impact velocity, a systematic increase in velocity spread toward the Sun and a transition in apsidal dependence: Earth and Mars exhibit nearly uniform distributions, while Venus and Mercury show strongly non-uniform behavior with enhanced encounters near $\varpi \approx 0^\circ$ and $180^\circ$, reflecting geometrically selective encounters in the inner solar system. These patterns reflect the transition from dynamically averaged encounters at larger heliocentric distances to geometrically selective encounters in the inner solar system. Low-eccentricity, low-inclination grains dominate the total flux, while a dynamically excited population of high-eccentricity, high-inclination grains controls the high-velocity tail. This two-population regime has direct implications for planetary processes: mass-driven processes such as regolith gardening are governed by the low-e population, while energy-driven processes such as impact vaporization and exosphere generation are dominated by a small fraction of high-velocity impacts. Together, these findings demonstrate that mean impact velocities are insufficient to characterize dust–planet interactions; the full velocity distribution must be considered when assessing impact-driven surface modification, exosphere generation, and volatile delivery across the terrestrial planets.

(d) Simplified dust flux models using power-law fits for all four inner planets (with R² > 0.92), offering simpler and observationally-grounded tools for flux estimation.

In addition, Spatial-temporal distribution of close encounters plots for flux estimation and Encounter probability ($P_0$) are provided. The spatial and temporal plots follow a clean quadratic scaling, confirming the geometric nature of dust-planet encounters in three-dimensional space. The temporal distribution exhibits a plateau phase and the duration of this plateau is size-dependent: smaller grains migrate faster due to stronger P-R drag and thus show shorter intervals, while larger grains remain longer in the terrestrial planet region. This size-dependent residence time directly influences the relative contribution of different grain sizes to the total impact flux. These plots reveal planet-specific asymmetries and encounter patterns previously undocumented. The fitted parameter $P_0$, which we tabulate for each planet, represents the encounter probability per unit area and serves as a reusable normalization factor for future studies.

The implications of our findings extend to planetary surface processes and mission planning. Despite its weak gravity, Mercury experiences high-velocity impacts ($\sim13.5$ km/s) that enhance vaporization and comminution, contributing to its heavily cratered surface and mature regolith. Mars experiences substantially gentler impacts ($\sim9.0$ km/s), reducing impact-induced melting and increasing the likelihood that fine-grained material and exogenous compounds, including organics, can survive emplacement at the Martian surface. The combined velocity distributions presented here provide essential input for quantitative models of impact vaporization, regolith gardening, and volatile processing. These velocity distributions, together with our calibrated fluxes, provide the necessary input for dust ablation and atmospheric entry models. Our calibrated fluxes offer a robust reference for comparing spacecraft observations from current and upcoming missions, including BepiColombo, MODEX, and VODEX. Future work will extend this framework to include cometary dust sources and incorporate collisional grooming to further refine our understanding of the inner solar system dust environment.

\begin{acknowledgments}
This work was carried out under the financial support of the INSPIRE Fellowship for Ph.D. provided by the Department of Science and Technology (DST), India. The computations were performed on the Param Vikram-1000 High Performance Computing Cluster of the Physical Research Laboratory, Ahmedabad (Department of Space), India. The authors would like to thank Prof. Francesco Marzari (University of Padua, Italy) for providing the basic dust dynamics code, which was adapted and extended for the present simulation work. We are also grateful to Dr. Deepak Jha for his valuable suggestions during the discussion. We also thank the reviewer for helpful suggestions.
\end{acknowledgments}

\begin{contribution}
Aanchal Sahu: Methodology, Investigation, Software, Data curation, Formal analysis, Validation, Visualization, Writing – original draft, Writing – review \& editing. 
Jayesh Pabari: Conceptualization, Validation, Supervision, Writing – review \& editing.
\end{contribution}

\bibliography{aanchal}{}

@article{Dermott,
 author = {Dermott, S. F. and Grogan, K. and Gustafson, B. A. S. and Jayaraman, S. and Kortenkamp, S. J. and Xu, Y. L.},
year = {1996},
month = {01},
pages = {143},
title = {Sources of Interplanetary Dust},
volume = {104},
journal = {International Astronomical Union Colloquium},
doi = {10.1017/S0252921100501444}
}

@article{GRUN1985244,
title = {Collisional balance of the meteoritic complex},
journal = {Icarus},
volume = {62},
number = {2},
pages = {244-272},
year = {1985},
issn = {0019-1035},
doi = {https://doi.org/10.1016/0019-1035(85)90121-6},
url = {https://www.sciencedirect.com/science/article/pii/0019103585901216},
author = {E. Grün and H.A. Zook and H. Fechtig and R.H. Giese}
}

@ARTICLE{1918AJ.....31..185H,
       author = {{Hirayama}, Kiyotsugu},
        title = "{Groups of asteroids probably of common origin}",
      journal = {\aj},
         year = 1918,
        month = oct,
       volume = {31},
        pages = {185-188},
          doi = {10.1086/104299},
       adsurl = {https://ui.adsabs.harvard.edu/abs/1918AJ.....31..185H}
}

@article{Dermott_Gomes_Durda_Gustafson_Jayaraman_Xu_Nicholson_1992, title={Dynamics of the Zodiacal Cloud}, volume={152}, DOI={10.1017/S007418090009135X}, journal={Symposium - International Astronomical Union}, author={Dermott, S. F. and Gomes, R. S. and Durda, D. D. and Gustafson, B. Å. S. and Jayaraman, S. and Xu, Y. L. and Nicholson, P. D.}, year={1992}, pages={333–347}}

@ARTICLE{2013Icar..226.1550K,
       author = {{Kortenkamp}, Stephen J.},
        title = "{Trapping and dynamical evolution of interplanetary dust particles in Earth{\textquoteright}s quasi-satellite resonance}",
      journal = {\icarus},
         year = 2013,
        month = nov,
       volume = {226},
       number = {2},
        pages = {1550-1558},
          doi = {10.1016/j.icarus.2013.08.020},
       adsurl = {https://ui.adsabs.harvard.edu/abs/2013Icar..226.1550K}
}

@ARTICLE{1990Icar...85..267S,
       author = {{Sykes}, M.~V.},
        title = "{Zodiacal dust bands: Their relation to asteroid families}",
      journal = {\icarus},
         year = 1990,
        month = jun,
       volume = {85},
       number = {2},
        pages = {267-289},
          doi = {10.1016/0019-1035(90)90117-R},
       adsurl = {https://ui.adsabs.harvard.edu/abs/1990Icar...85..267S}
}

@article{
doi:10.1126/science.262.5133.550,
author = {S. G. Love  and D. E. Brownlee },
title = {A Direct Measurement of the Terrestrial Mass Accretion Rate of Cosmic Dust},
journal = {Science},
volume = {262},
number = {5133},
pages = {550-553},
year = {1993},
doi = {10.1126/science.262.5133.550},
URL = {https://www.science.org/doi/abs/10.1126/science.262.5133.550},
eprint = {https://www.science.org/doi/pdf/10.1126/science.262.5133.550}}

@ARTICLE{1978A&A....67..381R,
       author = {{Roeser}, S. and {Staude}, H.~J.},
        title = "{The zodiacal light from 1500 {\r{A}} to 60 micron. Mie scattering and thermal emission.}",
      journal = {\aap},
         year = 1978,
        month = jul,
       volume = {67},
        pages = {381-394},
       adsurl = {https://ui.adsabs.harvard.edu/abs/1978A&A....67..381R}
}

@article{ZOOK1975183,
title = {A source for hyperbolic cosmic dust particles},
journal = {Planetary and Space Science},
volume = {23},
number = {1},
pages = {183-203},
year = {1975},
issn = {0032-0633},
doi = {https://doi.org/10.1016/0032-0633(75)90078-1},
url = {https://www.sciencedirect.com/science/article/pii/0032063375900781},
author = {Herbert A. Zook and Otto E. Berg}
}

@ARTICLE{1980P&SS...28..333G,
       author = {{Gruen}, E. and {Pailer}, N. and {Fechtig}, H. and {Kissel}, J.},
        title = "{Orbital and physical characteristics of micrometeoroids in the inner solar system as observed by Helios 1}",
      journal = {\planss},
         year = 1980,
        month = mar,
       volume = {28},
       number = {3},
        pages = {333-349},
          doi = {10.1016/0032-0633(80)90022-7},
       adsurl = {https://ui.adsabs.harvard.edu/abs/1980P&SS...28..333G}
}

@article{Borin2017,
author = {Borin, P. and Cremonese, Gabriele and Marzari, F. and Lucchetti, Alice},
year = {2017},
month = {06},
pages = {},
title = {Asteroidal and cometary dust flux in the inner solar system},
volume = {605},
journal = {Astronomy \& Astrophysics},
doi = {10.1051/0004-6361/201730617}
}

@ARTICLE{1950ApJ...111..134W,
       author = {{Wyatt}, S.~P. and {Whipple}, F.~L.},
        title = "{The Poynting-Robertson effect on meteor orbits}",
      journal = {\apj},
         year = 1950,
        month = jan,
       volume = {111},
        pages = {134-141},
          doi = {10.1086/145244},
       adsurl = {https://ui.adsabs.harvard.edu/abs/1950ApJ...111..134W}
}

@article{BURNS19791,
title = {Radiation forces on small particles in the solar system},
journal = {Icarus},
volume = {40},
number = {1},
pages = {1-48},
year = {1979},
issn = {0019-1035},
doi = {https://doi.org/10.1016/0019-1035(79)90050-2},
url = {https://www.sciencedirect.com/science/article/pii/0019103579900502},
author = {Joseph A. Burns and Philippe L. Lamy and Steven Soter}
}

@article{GUSTAFSON1987568,
title = {Interplanetary dust dynamics: II. Poynting-Robertson drag and planetary perturbations on cometary dust},
journal = {Icarus},
volume = {72},
number = {3},
pages = {568-581},
year = {1987},
issn = {0019-1035},
doi = {https://doi.org/10.1016/0019-1035(87)90053-4},
url = {https://www.sciencedirect.com/science/article/pii/0019103587900534},
author = {B.Å.S. Gustafson and N.Y. Misconi and E.T. Rusk}
}

@article{GUSTAFSON1986280,
title = {Interplanetary dust dynamics: I. Long-term gravitational effects of the inner planets on zodiacal dust},
journal = {Icarus},
volume = {66},
number = {2},
pages = {280-287},
year = {1986},
issn = {0019-1035},
doi = {https://doi.org/10.1016/0019-1035(86)90158-2},
url = {https://www.sciencedirect.com/science/article/pii/0019103586901582},
author = {Bo {\AA}.S. Gustafson and Nebil Y. Misconi}
}

@ARTICLE{1992Icar...97...70J,
       author = {{Jackson}, A.~A. and {Zook}, H.~A.},
        title = "{Orbital evolution of dust particles from comets and asteroids}",
      journal = {\icarus},
         year = 1992,
        month = may,
       volume = {97},
       number = {1},
        pages = {70-84},
          doi = {10.1016/0019-1035(92)90057-E},
       adsurl = {https://ui.adsabs.harvard.edu/abs/1992Icar...97...70J}
}

@INCOLLECTION{1990pihl.book..207L,
       author = {{Leinert}, C. and {Grun}, E.},
        title = "{Interplanetary Dust}",
    booktitle = {Physics of the Inner Heliosphere I},
         year = 1990,
       editor = {{Schwenn}, Rainer and {Marsch}, Eckart},
        pages = {207},
          doi = {10.1007/978-3-642-75361-9_5},
       adsurl = {https://ui.adsabs.harvard.edu/abs/1990pihl.book..207L}
}

@article{KORTENKAMP1998469,
title = {Accretion of Interplanetary Dust Particles by the Earth},
journal = {Icarus},
volume = {135},
number = {2},
pages = {469-495},
year = {1998},
issn = {0019-1035},
doi = {https://doi.org/10.1006/icar.1998.5994},
url = {https://www.sciencedirect.com/science/article/pii/S0019103598959942},
author = {Stephen J Kortenkamp and Stanley F Dermott}
}

@article{Molina2008article,
author = {Molina-Cuberos, Gregorio and Lopez-Moreno, J.-J and Arnold, F.},
year = {2008},
month = {06},
pages = {175-191},
title = {Meteoric Layers in Planetary Atmospheres},
volume = {137},
isbn = {978-0-387-87663-4},
journal = {Space Science Reviews},
doi = {10.1007/s11214-008-9340-5}
}

@article{PABARI2023105617,
title = {Metallic ion layers in planetary atmosphere: Boundary conditions and IDP flux},
journal = {Planetary and Space Science},
volume = {226},
pages = {105617},
year = {2023},
issn = {0032-0633},
doi = {https://doi.org/10.1016/j.pss.2022.105617},
url = {https://www.sciencedirect.com/science/article/pii/S0032063322002033},
author = {Jayesh P. Pabari and Srirag N. Nambiar and  Rashmi and Sonam Jitarwal}
}

@article{
doi:10.1126/science.1117755,
author = {M. Pätzold  and S. Tellmann  and B. Häusler  and D. Hinson  and R. Schaa  and G. L. Tyler },
title = {A Sporadic Third Layer in the Ionosphere of Mars},
journal = {Science},
volume = {310},
number = {5749},
pages = {837-839},
year = {2005},
doi = {10.1126/science.1117755},
URL = {https://www.science.org/doi/abs/10.1126/science.1117755},
eprint = {https://www.science.org/doi/pdf/10.1126/science.1117755}}

@article{HIRAI201487,
title = {Microparticle impact calibration of the Arrayed Large-Area Dust Detectors in INterplanetary space (ALADDIN) onboard the solar power sail demonstrator IKAROS},
journal = {Planetary and Space Science},
volume = {100},
pages = {87-97},
year = {2014},
note = {Cosmic Dust VI},
issn = {0032-0633},
doi = {https://doi.org/10.1016/j.pss.2014.05.009},
url = {https://www.sciencedirect.com/science/article/pii/S0032063314001366},
author = {Takayuki Hirai and Michael J. Cole and Masayuki Fujii and Sunao Hasegawa and Takeo Iwai and Masanori Kobayashi and Ralf Srama and Hajime Yano}
}

@article{33057761b02e4130b09a7d84297da8c2,
title = "Galileo Dust Detection System V4.1",
author = "H. Krueger and E. Gruen and M. Baguhl and D. Bindschadler and S. Dermott and N. Divine and H. Fechtig and A. Graps and B. Gustafson and D. Hamilton and M Hanner and M. Horanyi and J. Kissel and Lindblad, \{Bertil Anders\} and D. Linkert and G. Linkert and McDonnell, \{J. A. M.\} and I. Mann and R. Moissl and G. Morfill and C. Polanskey and G. Schwehm and R. Riemann and N. Siddique and R. Srama and P. Staubach and H. Zook",
year = "2010",
language = "English",
volume = "139",
journal = "NASA Planetary Data System",
publisher = "NASA",
}

@article{Leinert2007refId0,
	author = {{C. Leinert} and {B. Moster}},
	title = {Evidence for dust accumulation just outside the orbit of Venus},
	DOI= "10.1051/0004-6361:20077682",
	url= "https://doi.org/10.1051/0004-6361:20077682",
	journal = {A\&A},
	year = 2007,
	volume = 472,
	number = 1,
	pages = "335-340",
}

@article{
doi:10.1126/science.1243194,
author = {M. H. Jones  and D. Bewsher  and D. S. Brown },
title = {Imaging of a Circumsolar Dust Ring Near the Orbit of Venus},
journal = {Science},
volume = {342},
number = {6161},
pages = {960-963},
year = {2013},
doi = {10.1126/science.1243194},
URL = {https://www.science.org/doi/abs/10.1126/science.1243194},
eprint = {https://www.science.org/doi/pdf/10.1126/science.1243194}}

@article{JONES2017172,
title = {Mapping the circumsolar dust ring near the orbit of Venus},
journal = {Icarus},
volume = {288},
pages = {172-185},
year = {2017},
issn = {0019-1035},
doi = {https://doi.org/10.1016/j.icarus.2017.01.015},
url = {https://www.sciencedirect.com/science/article/pii/S001910351630392X},
author = {M.H. Jones and D. Bewsher and D.S. Brown}
}

@article{stenborg2021pristine,
  title={Pristine PSP/WISPR observations of the circumsolar dust ring near Venus's orbit},
  author={Stenborg, Guillermo and Gallagher, Brendan and Howard, Russell A and Hess, Phillip and Raouafi, Nour Eddine},
  journal={The Astrophysical Journal},
  volume={910},
  number={2},
  pages={157},
  year={2021},
  publisher={IOP Publishing}
}

@article{ISHIMOTO1996153,
title = {Formation of Phobos/Deimos Dust Rings},
journal = {Icarus},
volume = {122},
number = {1},
pages = {153-165},
year = {1996},
issn = {0019-1035},
doi = {https://doi.org/10.1006/icar.1996.0116},
url = {https://www.sciencedirect.com/science/article/pii/S0019103596901165},
author = {Hiroshi Ishimoto}
}

@article{PABARI20171,
title = {Estimation of micrometeorites and satellite dust flux surrounding Mars in the light of MAVEN results},
journal = {Icarus},
volume = {288},
pages = {1-9},
year = {2017},
issn = {0019-1035},
doi = {https://doi.org/10.1016/j.icarus.2017.01.023},
url = {https://www.sciencedirect.com/science/article/pii/S0019103516303311},
author = {J.P. Pabari and P.J. Bhalodi}
}

@article{https://doi.org/10.1029/2020JE006509,
author = {Jorgensen, J. L. and Benn, M. and Connerney, J. E. P. and Denver, T. and Jorgensen, P. S. and Andersen, A. C. and Bolton, S. J.},
title = {Distribution of Interplanetary Dust Detected by the Juno Spacecraft and Its Contribution to the Zodiacal Light},
journal = {Journal of Geophysical Research: Planets},
volume = {126},
number = {3},
pages = {e2020JE006509},
doi = {https://doi.org/10.1029/2020JE006509},
url = {https://agupubs.onlinelibrary.wiley.com/doi/abs/10.1029/2020JE006509},
eprint = {https://agupubs.onlinelibrary.wiley.com/doi/pdf/10.1029/2020JE006509},
note = {e2020JE006509 2020JE006509},
year = {2021}
}

@article{10.1093/mnras/stad1045,
    author = {Pabari, J P},
    title = {Likelihood of Martian moons as dust sources in light with Juno observations},
    journal = {Monthly Notices of the Royal Astronomical Society},
    volume = {522},
    number = {1},
    pages = {1428-1440},
    year = {2023},
    month = {04},
    issn = {0035-8711},
    doi = {10.1093/mnras/stad1045},
    url = {https://doi.org/10.1093/mnras/stad1045},
    eprint = {https://academic.oup.com/mnras/article-pdf/522/1/1428/50048738/stad1045.pdf},
}

@ARTICLE{2002M&PSA..37R.126S,
       author = {{Sasaki}, S. and {Igenbergs}, E. and {Ohashi}, H. and {Senger}, R. and {Hofschuster}, G. and {M{\"u}nzenmayer}, R. and {Naumann}, W. and {Gr{\"u}n}, E. and {Hamabe}, Y. and {Mann}, I. and {Nogami}, K. and {Svedhem}, H.},
        title = "{Four-Year Observation of Interplanetary and Interstellar Dust by NOZOMI-MDC (Mars Dust Counter)}",
      journal = {Meteoritics and Planetary Science Supplement},
         year = 2002,
        month = jul,
       volume = {37},
        pages = {A126},
       adsurl = {https://ui.adsabs.harvard.edu/abs/2002M&PSA..37R.126S}
}

@ARTICLE{1990JGR....9514497F,
       author = {{Flynn}, George J. and {McKay}, David S.},
        title = "{An assessment of the meteoritic contribution to the Martian soil.}",
      journal = {\jgr},
         year = 1990,
        month = aug,
       volume = {95},
        pages = {14497-14509},
          doi = {10.1029/JB095iB09p14497},
       adsurl = {https://ui.adsabs.harvard.edu/abs/1990JGR....9514497F}
}

@ARTICLE{1994A&A...283..275M,
       author = {{Marzari}, F. and {Vanzani}, V.},
        title = "{Dynamical evolution of interplanetary dust particles}",
      journal = {\aap},
         year = 1994,
        month = mar,
       volume = {283},
       number = {1},
        pages = {275-286},
       adsurl = {https://ui.adsabs.harvard.edu/abs/1994A&A...283..275M}
}

@article{Sykes1986,
  author = {Sykes, M. V. and Greenberg, R.},
  title = {Asteroid rotation rates depend on size and type},
  journal = {Icarus},
  year = {1986},
  volume = {65},
  pages = {51-72},
  doi = {10.1016/0019-1035(86)90062-1}
}

@INPROCEEDINGS{1989aste.conf..316G,
       author = {{Gradie}, Jonathan C. and {Chapman}, Clark R. and {Tedesco}, Edward F.},
        title = "{Distribution of taxonomic classes and the compositional structure of the asteroid belt.}",
    booktitle = {Asteroids II},
         year = 1989,
       editor = {{Binzel}, Richard P. and {Gehrels}, Tom and {Matthews}, Mildred Shapley},
        month = jan,
        pages = {316-335},
       adsurl = {https://ui.adsabs.harvard.edu/abs/1989aste.conf..316G}
}

@article{Milani1994,
  author = {Milani, A. and Knežević, Z.},
  title = {Asteroid proper elements and the dynamical structure of the asteroid main belt},
  journal = {Icarus},
  year = {1994},
  volume = {107},
  pages = {219-254},
  doi = {10.1006/icar.1994.1021}
}

@article{Sykes2004,
  author = {Sykes, M. V. and Grün, E. and Reach, W. T. and Jenniskens, P.},
  title = {The interplanetary dust complex and comets},
  journal = {Comets II},
  year = {2004},
  editor = {Festou, M. C. and Keller, H. U. and Weaver, H. A.},
  pages = {677-693},
  publisher = {University of Arizona Press}
}

@inproceedings{Pabari2016MARSOD,
  author    = {Pabari, Jayesh P. and Bhalodi, P. J. and Patel, D. K.},
  title     = {Mars Orbit Dust Experiment (MODEX) for Future Mars Orbiter},
  booktitle = {Proceedings of the 47th Lunar and Planetary Science Conference},
  year      = {2016},
}

@article{MARZARI1996192,
title = {Collision Rates and Impact Velocities in the Trojan Asteroid Swarms},
journal = {Icarus},
volume = {119},
number = {1},
pages = {192-201},
year = {1996},
issn = {0019-1035},
doi = {https://doi.org/10.1006/icar.1996.0011},
url = {https://www.sciencedirect.com/science/article/pii/S0019103596900111},
author = {{Marzari}, F. and {Scholl}, H. and {Farinella}, P.}
}

@article{Everhart1974,
  author  = {E. Everhart},
  title   = {Implicit single-sequence methods for integrating orbits},
  journal = {Celestial Mechanics},
  year    = {1974},
  volume  = {10},
  number  = {1},
  pages   = {35--55},
  doi     = {10.1007/BF01261877},
  url     = {https://doi.org/10.1007/BF01261877}
}

@article{withers,
author = {Withers, Paul and Mendillo, Michael and Hinson, David and Cahoy, Kerri},
year = {2008},
month = {12},
pages = {},
title = {Physical characteristics and occurrence rates of meteoric plasma layers detected in the Martian ionosphere by Mars Global Surveyor Radio Science Experiment},
volume = {113},
journal = {Journal of Geophysical Research},
doi = {10.1029/2008JA013636}
}

@Article{Mukai:1984:MSD,
  author = {{T. Mukai} and {R. H. Giese}},
  title = {{Modification of the spatial distribution of interplanetary dust grains by Lorentz forces}},
  journal = {{Astronomy and Astrophysics}},
  volume = {131},
  pages = {355-363},
  year = {1984},
}

@ARTICLE{2016A&A...588C...3B,
       author = {{Borin}, P. and {Cremonese}, G. and {Marzari}, F.},
        title = "{Statistical analysis of the flux of micrometeoroids at Mercury from both cometary and asteroidal components (Corrigendum)}",
      journal = {\aap},
         year = 2016,
        month = apr,
       volume = {588},
          eid = {C3},
        pages = {C3},
          doi = {10.1051/0004-6361/201526767e},
       adsurl = {https://ui.adsabs.harvard.edu/abs/2016A&A...588C...3B}
}

@ARTICLE{2009A&A...503..259B,
       author = {{Borin}, P. and {Cremonese}, G. and {Marzari}, F. and {Bruno}, M. and {Marchi}, S.},
        title = "{Statistical analysis of micrometeoroids flux on Mercury}",
      journal = {\aap},
         year = 2009,
        month = aug,
       volume = {503},
       number = {1},
        pages = {259-264},
          doi = {10.1051/0004-6361/200912080},
       adsurl = {https://ui.adsabs.harvard.edu/abs/2009A&A...503..259B}
}

@ARTICLE{1982A&A...107...97M,
       author = {{Mukai}, T. and {Yamamoto}, T.},
        title = "{Solar wind pressure on interplanetary dust}",
      journal = {\aap},
         year = 1982,
        month = mar,
       volume = {107},
       number = {1},
        pages = {97-100},
       adsurl = {https://ui.adsabs.harvard.edu/abs/1982A&A...107...97M}
}

@book{roy2004orbital,
  title={Orbital Motion},
  author={Roy, A. E.},
  year={2004},
  edition={4th},
  publisher={CRC Press},
  doi={10.1201/9780367806620},
  url={https://doi.org/10.1201/9780367806620}
}

@book{Fitzpatrick_2012, place={Cambridge}, title={An Introduction to Celestial Mechanics}, publisher={Cambridge University Press}, author={Fitzpatrick, Richard}, year={2012}}

@article{MOLINACUBEROS2001143,
title = {Ionospheric layer induced by meteoric ionization in Titan's atmosphere},
journal = {Planetary and Space Science},
volume = {49},
number = {2},
pages = {143-153},
year = {2001},
issn = {0032-0633},
doi = {https://doi.org/10.1016/S0032-0633(00)00133-1},
url = {https://www.sciencedirect.com/science/article/pii/S0032063300001331},
author = {G.J Molina-Cuberos and H Lammer and W Stumptner and K Schwingenschuh and H.O Rucker and J.J López-Moreno and R Rodrigo and T Tokano}
}

@INPROCEEDINGS{Everhart1985,
       author = {{E. Everhart}},
        title = "{An efficient integrator that uses Gauss-Radau spacings}",
    booktitle = {IAU Colloq. 83: Dynamics of Comets: Their Origin and Evolution},
         year = 1985,
       editor = {{Carusi}, A. and {Valsecchi}, G.~B.},
       series = {Astrophysics and Space Science Library},
       volume = {115},
        month = jan,
        pages = {185},
          doi = {10.1007/978-94-009-5400-7_17},
       adsurl = {https://ui.adsabs.harvard.edu/abs/1985ASSL..115..185E}
}

@article{Hill0eb94dbb-e87d-35f4-a82b-f98b26a6e529,
 ISSN = {00029327, 10806377},
 URL = {http://www.jstor.org/stable/2369430},
 author = {G. W. Hill},
 journal = {American Journal of Mathematics},
 number = {1},
 pages = {5--26},
 publisher = {Johns Hopkins University Press},
 title = {Researches in the Lunar Theory},
 urldate = {2025-12-24},
 volume = {1},
 year = {1878}
}

@article{Pabari2025DEX,
  author  = {Pabari, Jayesh and
             Nambiar, Srirag and
             Singh, Rashmi and
             Jitarwal, Sonam and
             Bhardwaj, Anil and
             Praneeth, S. M. K. and
             Shah, Bhavik and
             Suthar, Pinal and
             Pandya, Shilpa and
             Rami, Jaimin and
             Kumar, Deepak and
             Singh, V. K. and
             Khandekar, Rahul and
             Singh, Rajesh Kumar and
             Sahu, Aanchal and
             Adalja, Hiteshkumr and
             Patel, Arpit},
  title   = {DEX in near Earth orbit in light of {V}enus {O}rbiter {D}ust {E}xperiment},
  journal = {Scientific Reports},
  year    = {2025},
  volume  = {15},
  number  = {1},
  pages   = {38168},
  doi     = {10.1038/s41598-025-21988-2},
}

@ARTICLE{2012ApJ...749L..40C,
       author = {{Cremonese}, G. and {Borin}, P. and {Martellato}, E. and {Marzari}, F. and {Bruno}, M.},
        title = "{New Calibration of the Micrometeoroid Flux on Earth}",
      journal = {\apjl},
         year = 2012,
        month = apr,
       volume = {749},
       number = {2},
          eid = {L40},
        pages = {L40},
          doi = {10.1088/2041-8205/749/2/L40},
       adsurl = {https://ui.adsabs.harvard.edu/abs/2012ApJ...749L..40C}
}

@ARTICLE{Andersson2015Sci...350.0398A,
       author = {{Andersson}, L. and {Weber}, T.~D. and {Malaspina}, D. and {Crary}, F. and {Ergun}, R.~E. and {Delory}, G.~T. and {Fowler}, C.~M. and {Morooka}, M.~W. and {McEnulty}, T. and {Eriksson}, A.~I. and {Andrews}, D.~J. and {Horanyi}, M. and {Collette}, A. and {Yelle}, R. and {Jakosky}, B.~M.},
        title = "{Dust observations at orbital altitudes surrounding Mars}",
      journal = {Science},
         year = 2015,
        month = nov,
       volume = {350},
       number = {6261},
          eid = {0398},
        pages = {0398},
          doi = {10.1126/science.aad0398},
       adsurl = {https://ui.adsabs.harvard.edu/abs/2015Sci...350.0398A}
}

@article{PABARI201868,
title = {Orbital altitude dust at Mars, its implication and a prototype for its detection},
journal = {Planetary and Space Science},
volume = {161},
pages = {68-75},
year = {2018},
issn = {0032-0633},
doi = {https://doi.org/10.1016/j.pss.2018.06.008},
url = {https://www.sciencedirect.com/science/article/pii/S0032063317304750},
author = {J.P. Pabari and S.A. Haider and B.M. Pandya and R.K. Singh and A. Kumar and D.K. Patel and A. Bogavelly}
}

@article{CARRILLOSANCHEZ2020113395,
title = {Cosmic dust fluxes in the atmospheres of Earth, Mars, and Venus},
journal = {Icarus},
volume = {335},
pages = {113395},
year = {2020},
issn = {0019-1035},
doi = {https://doi.org/10.1016/j.icarus.2019.113395},
url = {https://www.sciencedirect.com/science/article/pii/S0019103519301824},
author = {Juan Diego Carrillo-Sánchez and Juan Carlos Gómez-Martín and David L. Bones and David Nesvorný and Petr Pokorný and Mehdi Benna and George J. Flynn and John M.C. Plane}
}

@ARTICLE{Cintala1992JGR....97..947C,
       author = {{Cintala}, Mark J.},
        title = "{Impact-induced thermal effects in the lunar and mercurian regoliths}",
      journal = {\jgr},
         year = 1992,
        month = jan,
       volume = {97},
       number = {E1},
        pages = {947-973},
          doi = {10.1029/91JE02207},
       adsurl = {https://ui.adsabs.harvard.edu/abs/1992JGR....97..947C}
}

@article{MULLER20021101,
title = {Estimation of the dust flux near Mercury},
journal = {Planetary and Space Science},
volume = {50},
number = {10},
pages = {1101-1115},
year = {2002},
issn = {0032-0633},
doi = {https://doi.org/10.1016/S0032-0633(02)00048-X},
url = {https://www.sciencedirect.com/science/article/pii/S003206330200048X},
author = {M Müller and S.F Green and N McBride and D Koschny and J.C Zarnecki and M.S Bentley}
}

@article{BATYGIN20191,
title = {The planet nine hypothesis},
journal = {Physics Reports},
volume = {805},
pages = {1-53},
year = {2019},
note = {The planet nine hypothesis},
issn = {0370-1573},
doi = {https://doi.org/10.1016/j.physrep.2019.01.009},
url = {https://www.sciencedirect.com/science/article/pii/S037015731930047X},
author = {Konstantin Batygin and Fred C. Adams and Michael E. Brown and Juliette C. Becker}
}

@article{Mannarticle,
author = {Mann, Ingrid and Czechowski, A.},
year = {2020},
month = {12},
pages = {},
title = {Dust observations from Parker Solar Probe: Dust ejection from the inner Solar System},
volume = {650},
journal = {Astronomy \& Astrophysics},
doi = {10.1051/0004-6361/202039362}
}

@article{NESVORNY2026116768,
title = {Discovery of 63 new young asteroid families},
journal = {Icarus},
volume = {443},
pages = {116768},
year = {2026},
issn = {0019-1035},
doi = {https://doi.org/10.1016/j.icarus.2025.116768},
url = {https://www.sciencedirect.com/science/article/pii/S0019103525003161},
author = {David Nesvorný and David Vokrouhlický and Miroslav Brož and Fernando V. Roig}
}

@ARTICLE{2010ApJ...713..816N,
       author = {{Nesvorn{\'y}}, David and {Jenniskens}, Peter and {Levison}, Harold F. and {Bottke}, William F. and {Vokrouhlick{\'y}}, David and {Gounelle}, Matthieu},
        title = "{Cometary Origin of the Zodiacal Cloud and Carbonaceous Micrometeorites. Implications for Hot Debris Disks}",
      journal = {\apj},
         year = 2010,
        month = apr,
       volume = {713},
       number = {2},
        pages = {816-836},
          doi = {10.1088/0004-637X/713/2/816},
archivePrefix = {arXiv},
       eprint = {0909.4322},
 primaryClass = {astro-ph.EP},
       adsurl = {https://ui.adsabs.harvard.edu/abs/2010ApJ...713..816N}
}

@article{Pokornyarticle,
author = {Pokorný, Petr and Brown, Peter},
year = {2016},
month = {05},
pages = {},
title = {A reproducible method to determine the meteoroid mass index},
volume = {592},
journal = {Astronomy \& Astrophysics},
doi = {10.1051/0004-6361/201628134}
}

@ARTICLE{Ade2014A&A...571A..14P,
       author = {{Planck Collaboration} and {Ade}, P.~A.~R. and {Aghanim}, N. and {Armitage-Caplan}, C. and {Arnaud}, M. and {Ashdown}, M. and {Atrio-Barandela}, F. and {Aumont}, J. and {Baccigalupi}, C. and {Banday}, A.~J. and {Barreiro}, R.~B. and {Bartlett}, J.~G. and {Battaner}, E. and {Benabed}, K. and {Beno{\^\i}t}, A. and {Benoit-L{\'e}vy}, A. and {Bernard}, J.-P. and {Bersanelli}, M. and {Bielewicz}, P. and {Bobin}, J. and {Bock}, J.~J. and {Bonaldi}, A. and {Bond}, J.~R. and {Borrill}, J. and {Bouchet}, F.~R. and {Boulanger}, F. and {Bridges}, M. and {Bucher}, M. and {Burigana}, C. and {Butler}, R.~C. and {Cardoso}, J.-F. and {Catalano}, A. and {Chamballu}, A. and {Chary}, R.-R. and {Chen}, X. and {Chiang}, H.~C. and {Chiang}, L.-Y. and {Christensen}, P.~R. and {Church}, S. and {Clements}, D.~L. and {Colley}, J.-M. and {Colombi}, S. and {Colombo}, L.~P.~L. and {Couchot}, F. and {Coulais}, A. and {Crill}, B.~P. and {Curto}, A. and {Cuttaia}, F. and {Danese}, L. and {Davies}, R.~D. and {de Bernardis}, P. and {de Rosa}, A. and {de Zotti}, G. and {Delabrouille}, J. and {Delouis}, J.-M. and {D{\'e}sert}, F.-X. and {Dickinson}, C. and {Diego}, J.~M. and {Dole}, H. and {Donzelli}, S. and {Dor{\'e}}, O. and {Douspis}, M. and {Dupac}, X. and {Efstathiou}, G. and {En{\ss}lin}, T.~A. and {Eriksen}, H.~K. and {Finelli}, F. and {Forni}, O. and {Frailis}, M. and {Fraisse}, A.~A. and {Franceschi}, E. and {Galeotta}, S. and {Ganga}, K. and {Giard}, M. and {Giraud-H{\'e}raud}, Y. and {Gonz{\'a}lez-Nuevo}, J. and {G{\'o}rski}, K.~M. and {Gratton}, S. and {Gregorio}, A. and {Gruppuso}, A. and {Hansen}, F.~K. and {Hanson}, D. and {Harrison}, D. and {Helou}, G. and {Henrot-Versill{\'e}}, S. and {Hern{\'a}ndez-Monteagudo}, C. and {Herranz}, D. and {Hildebrandt}, S.~R. and {Hivon}, E. and {Hobson}, M. and {Holmes}, W.~A. and {Hornstrup}, A. and {Hovest}, W. and {Huffenberger}, K.~M. and {Jaffe}, A.~H. and {Jaffe}, T.~R. and {Jones}, W.~C. and {Juvela}, M. and {Keih{\"a}nen}, E. and {Keskitalo}, R. and {Kisner}, T.~S. and {Kneissl}, R. and {Knoche}, J. and {Knox}, L. and {Kunz}, M. and {Kurki-Suonio}, H. and {Lagache}, G. and {L{\"a}hteenm{\"a}ki}, A. and {Lamarre}, J.-M. and {Lasenby}, A. and {Laureijs}, R.~J. and {Lawrence}, C.~R. and {Leonardi}, R. and {Lesgourgues}, J. and {Liguori}, M. and {Lilje}, P.~B. and {Linden-V{\o}rnle}, M. and {L{\'o}pez-Caniego}, M. and {Lubin}, P.~M. and {Mac{\'\i}as-P{\'e}rez}, J.~F. and {Maffei}, B. and {Maino}, D. and {Mandolesi}, N. and {Maris}, M. and {Marshall}, D.~J. and {Martin}, P.~G. and {Mart{\'\i}nez-Gonz{\'a}lez}, E. and {Masi}, S. and {Massardi}, M. and {Matarrese}, S. and {Matthai}, F. and {Mazzotta}, P. and {Meinhold}, P.~R. and {Melchiorri}, A. and {Mendes}, L. and {Mennella}, A. and {Migliaccio}, M. and {Mitra}, S. and {Miville-Desch{\^e}nes}, M.-A. and {Moneti}, A. and {Montier}, L. and {Morgante}, G. and {Mortlock}, D. and {Mottet}, S. and {Munshi}, D. and {Murphy}, J.~A. and {Naselsky}, P. and {Nati}, F. and {Natoli}, P. and {Netterfield}, C.~B. and {N{\o}rgaard-Nielsen}, H.~U. and {Noviello}, F. and {Novikov}, D. and {Novikov}, I. and {Osborne}, S. and {O'Sullivan}, C. and {Oxborrow}, C.~A. and {Paci}, F. and {Pagano}, L. and {Pajot}, F. and {Paladini}, R. and {Paoletti}, D. and {Pasian}, F. and {Patanchon}, G. and {Perdereau}, O. and {Perotto}, L. and {Perrotta}, F. and {Piacentini}, F. and {Piat}, M. and {Pierpaoli}, E. and {Pietrobon}, D. and {Plaszczynski}, S. and {Pointecouteau}, E. and {Polegre}, A.~M. and {Polenta}, G. and {Ponthieu}, N. and {Popa}, L. and {Poutanen}, T. and {Pratt}, G.~W. and {Pr{\'e}zeau}, G. and {Prunet}, S. and {Puget}, J.-L. and {Rachen}, J.~P. and {Reach}, W.~T. and {Rebolo}, R. and {Reinecke}, M. and {Remazeilles}, M. and {Renault}, C. and {Ricciardi}, S. and {Riller}, T. and {Ristorcelli}, I. and {Rocha}, G. and {Rosset}, C. and {Roudier}, G. and {Rowan-Robinson}, M. and {Rusholme}, B. and {Sandri}, M. and {Santos}, D. and {Savini}, G. and {Scott}, D.},
        title = "{Planck 2013 results. XIV. Zodiacal emission}",
      journal = {\aap},
         year = 2014,
        month = nov,
       volume = {571},
          eid = {A14},
        pages = {A14},
          doi = {10.1051/0004-6361/201321562},
archivePrefix = {arXiv},
       eprint = {1303.5074},
 primaryClass = {astro-ph.CO},
       adsurl = {https://ui.adsabs.harvard.edu/abs/2014A&A...571A..14P}
}

@ARTICLE{Szalay2024PSJ.....5..266S,
       author = {{Szalay}, J.~R. and {Pokorn{\'y}}, P. and {Malaspina}, D.~M.},
        title = "{Size Distribution of Small Grains in the Inner Zodiacal Cloud}",
      journal = {\psj},
         year = 2024,
        month = dec,
       volume = {5},
       number = {12},
          eid = {266},
        pages = {266},
          doi = {10.3847/PSJ/ad8b27},
archivePrefix = {arXiv},
       eprint = {2409.07411},
 primaryClass = {astro-ph.EP},
       adsurl = {https://ui.adsabs.harvard.edu/abs/2024PSJ.....5..266S}
}

@ARTICLE{2024PSJ.....5...82P,
       author = {{Pokorn{\'y}}, Petr and {Moorhead}, Althea V. and {Kuchner}, Marc J. and {Szalay}, Jamey R. and {Malaspina}, David M.},
        title = "{How Long-lived Grains Dominate the Shape of the Zodiacal Cloud}",
      journal = {\psj},
         year = 2024,
        month = mar,
       volume = {5},
       number = {3},
          eid = {82},
        pages = {82},
          doi = {10.3847/PSJ/ad2de8},
archivePrefix = {arXiv},
       eprint = {2401.13776},
 primaryClass = {astro-ph.EP},
       adsurl = {https://ui.adsabs.harvard.edu/abs/2024PSJ.....5...82P}
}

@article{WILCK1996493,
title = {Radiation pressure forces on “typical” interplanetary dust grains},
journal = {Planetary and Space Science},
volume = {44},
number = {5},
pages = {493-499},
year = {1996},
issn = {0032-0633},
doi = {https://doi.org/10.1016/0032-0633(95)00151-4},
url = {https://www.sciencedirect.com/science/article/pii/0032063395001514},
author = {M. Wilck and I. Mann}
}

@ARTICLE{2020ApJ...892..115M,
       author = {{Malaspina}, David M. and {Szalay}, Jamey R. and {Pokorn{\'y}}, Petr and {Page}, Brent and {Bale}, Stuart D. and {Bonnell}, John W. and {de Wit}, Thierry Dudok and {Goetz}, Keith and {Goodrich}, Katherine and {Harvey}, Peter R. and {MacDowall}, Robert J. and {Pulupa}, Marc},
        title = "{In Situ Observations of Interplanetary Dust Variability in the Inner Heliosphere}",
      journal = {\apj},
         year = 2020,
        month = apr,
       volume = {892},
       number = {2},
          eid = {115},
        pages = {115},
          doi = {10.3847/1538-4357/ab799b},
       adsurl = {https://ui.adsabs.harvard.edu/abs/2020ApJ...892..115M}
}

@ARTICLE{2024ApJ...972...24S,
       author = {{Stenborg}, Guillermo and {Vourlidas}, Angelos and {Paouris}, Evangelos and {Howard}, Russell A.},
        title = "{Novel Insights on the Dust Distribution in the Zodiacal Dust Cloud from PSP/WISPR Observations at Large Elongations}",
      journal = {\apj},
         year = 2024,
        month = sep,
       volume = {972},
       number = {1},
          eid = {24},
        pages = {24},
          doi = {10.3847/1538-4357/ad58b6},
       adsurl = {https://ui.adsabs.harvard.edu/abs/2024ApJ...972...24S}
}

@ARTICLE{2021PSJ.....2..185S,
       author = {{Szalay}, J.~R. and {Pokorn{\'y}}, P. and {Malaspina}, D.~M. and {Pusack}, A. and {Bale}, S.~D. and {Battams}, K. and {Gasque}, L.~C. and {Goetz}, K. and {Kr{\"u}ger}, H. and {McComas}, D.~J. and {Schwadron}, N.~A. and {Strub}, P.},
        title = "{Collisional Evolution of the Inner Zodiacal Cloud}",
      journal = {\psj},
         year = 2021,
        month = oct,
       volume = {2},
       number = {5},
          eid = {185},
        pages = {185},
          doi = {10.3847/PSJ/abf928},
archivePrefix = {arXiv},
       eprint = {2104.08217},
 primaryClass = {astro-ph.EP},
       adsurl = {https://ui.adsabs.harvard.edu/abs/2021PSJ.....2..185S}
}

@ARTICLE{2011ApJ...743..129N,
       author = {{Nesvorn{\'y}}, David and {Janches}, Diego and {Vokrouhlick{\'y}}, David and {Pokorn{\'y}}, Petr and {Bottke}, William F. and {Jenniskens}, Peter},
        title = "{Dynamical Model for the Zodiacal Cloud and Sporadic Meteors}",
      journal = {\apj},
         year = 2011,
        month = dec,
       volume = {743},
       number = {2},
          eid = {129},
        pages = {129},
          doi = {10.1088/0004-637X/743/2/129},
archivePrefix = {arXiv},
       eprint = {1109.2983},
 primaryClass = {astro-ph.EP},
       adsurl = {https://ui.adsabs.harvard.edu/abs/2011ApJ...743..129N}
}

@ARTICLE{2018ApJ...863...31P,
       author = {{Pokorn{\'y}}, Petr and {Sarantos}, Menelaos and {Janches}, Diego},
        title = "{A Comprehensive Model of the Meteoroid Environment around Mercury}",
      journal = {\apj},
         year = 2018,
        month = aug,
       volume = {863},
       number = {1},
          eid = {31},
        pages = {31},
          doi = {10.3847/1538-4357/aad051},
archivePrefix = {arXiv},
       eprint = {1807.02749},
 primaryClass = {astro-ph.EP},
       adsurl = {https://ui.adsabs.harvard.edu/abs/2018ApJ...863...31P}
}

@ARTICLE{2022PSJ.....3..239C,
       author = {{Carrillo-S{\'a}nchez}, Juan Diego and {Janches}, Diego and {Plane}, John M.~C. and {Pokorn{\'y}}, Petr and {Sarantos}, Menelaos and {Crismani}, Matteo M.~J. and {Feng}, Wuhu and {Marsh}, Daniel R.},
        title = "{A Modeling Study of the Seasonal, Latitudinal, and Temporal Distribution of the Meteoroid Mass Input at Mars: Constraining the Deposition of Meteoric Ablated Metals in the Upper Atmosphere}",
      journal = {\psj},
         year = 2022,
        month = oct,
       volume = {3},
       number = {10},
          eid = {239},
        pages = {239},
          doi = {10.3847/PSJ/ac8540},
       adsurl = {https://ui.adsabs.harvard.edu/abs/2022PSJ.....3..239C}
}

@article{Pokorný_2019,
doi = {10.3847/2041-8213/ab0827},
url = {https://doi.org/10.3847/2041-8213/ab0827},
year = {2019},
month = {mar},
publisher = {The American Astronomical Society},
volume = {873},
number = {2},
pages = {L16},
author = {Pokorný, Petr and Kuchner, Marc},
title = {Co-orbital Asteroids as the Source of Venus's Zodiacal Dust Ring},
journal = {The Astrophysical Journal Letters}
}

@article{Stenborg_2018,
doi = {10.3847/1538-4357/aae6cb},
url = {https://doi.org/10.3847/1538-4357/aae6cb},
year = {2018},
month = {nov},
publisher = {The American Astronomical Society},
volume = {868},
number = {1},
pages = {74},
author = {Stenborg, Guillermo and Stauffer, Johnathan R. and Howard, Russell A.},
title = {Evidence for a Circumsolar Dust Ring Near Mercury’s Orbit},
journal = {The Astrophysical Journal}
}

@article{Pokorný_2023,
doi = {10.3847/PSJ/acb52e},
url = {https://doi.org/10.3847/PSJ/acb52e},
year = {2023},
month = {feb},
publisher = {The American Astronomical Society},
volume = {4},
number = {2},
pages = {33},
author = {Pokorný, Petr and Deutsch, Ariel N. and Kuchner, Marc J.},
title = {Mercury's Circumsolar Dust Ring as an Imprint of a Recent Impact},
journal = {The Planetary Science Journal}
}

@article{Pokorný_2022,
doi = {10.3847/PSJ/ac4019},
url = {https://doi.org/10.3847/PSJ/ac4019},
year = {2022},
month = {jan},
publisher = {The American Astronomical Society},
volume = {3},
number = {1},
pages = {14},
author = {Pokorný, Petr and Szalay, Jamey R. and Horányi, Mihály and Kuchner, Marc J.},
title = {Modeling Meteoroid Impacts on the Juno Spacecraft},
journal = {The Planetary Science Journal}
}

@article{KOBAYASHI201841,
title = {In situ observations of dust particles in Martian dust belts using a large-sensitive-area dust sensor},
journal = {Planetary and Space Science},
volume = {156},
pages = {41-46},
year = {2018},
note = {Dust, Atmosphere, and Plasma Environment of the Moon and Small Bodies},
issn = {0032-0633},
doi = {https://doi.org/10.1016/j.pss.2017.12.011},
url = {https://www.sciencedirect.com/science/article/pii/S0032063317302891},
author = {Masanori Kobayashi and Harald Krüger and Hiroki Senshu and Koji Wada and Osamu Okudaira and Sho Sasaki and Hiroshi Kimura}
}

@ARTICLE{1951PRIA...54..165O,
       author = {{{\"O}pik}, E.~J.},
        title = "{Collision probabilities with the planets and the distribution of interplanetary matter}",
      journal = {Pattern Recognition and Image Analysis},
         year = 1951,
        month = jan,
       volume = {54},
        pages = {165-199},
       adsurl = {https://ui.adsabs.harvard.edu/abs/1951PRIA...54..165O}
}

@article{Wetherill1967CollisionsIT,
  title={Collisions in the asteroid belt},
  author={George W. Wetherill},
  journal={Journal of Geophysical Research},
  year={1967},
  volume={72},
  pages={2429-2444},
  url={https://api.semanticscholar.org/CorpusID:119416963}
}

@article{POKORNY2013682,
title = {Öpik-type collision probability for high-inclination orbits: Targets on eccentric orbits},
journal = {Icarus},
volume = {226},
number = {1},
pages = {682-693},
year = {2013},
issn = {0019-1035},
doi = {https://doi.org/10.1016/j.icarus.2013.06.015},
url = {https://www.sciencedirect.com/science/article/pii/S0019103513002716},
author = {Petr Pokorný and David Vokrouhlický}
}

@ARTICLE{2006A&A...455..509K,
       author = {{Krivov}, A.~V. and {L{\"o}hne}, T. and {Srem{\v{c}}evi{\'c}}, M.},
        title = "{Dust distributions in debris disks: effects of gravity, radiation pressure and collisions}",
      journal = {\aap},
         year = 2006,
        month = aug,
       volume = {455},
       number = {2},
        pages = {509-519},
          doi = {10.1051/0004-6361:20064907},
       adsurl = {https://ui.adsabs.harvard.edu/abs/2006A&A...455..509K}
}

@ARTICLE{Sommer2020A&A...635A..10S,
       author = {{Sommer}, M. and {Yano}, H. and {Srama}, R.},
        title = "{Effects of neighbouring planets on the formation of resonant dust rings in the inner Solar System}",
      journal = {\aap},
         year = 2020,
        month = mar,
       volume = {635},
          eid = {A10},
        pages = {A10},
          doi = {10.1051/0004-6361/201936676},
archivePrefix = {arXiv},
       eprint = {2001.07611},
 primaryClass = {astro-ph.EP},
       adsurl = {https://ui.adsabs.harvard.edu/abs/2020A&A...635A..10S}
}

@article{Pokorny2018article,
author = {Pokorný, Petr and Sarantos, M. and Janches, Diego},
year = {2018},
month = {08},
pages = {31},
title = {A Comprehensive Model of the Meteoroid Environment around Mercury},
volume = {863},
journal = {The Astrophysical Journal},
doi = {10.3847/1538-4357/aad051}
}

@article{Thorpe_2019,
doi = {10.3847/1538-4357/ab3649},
url = {https://doi.org/10.3847/1538-4357/ab3649},
year = {2019},
month = {sep},
publisher = {The American Astronomical Society},
volume = {883},
number = {1},
pages = {53},
author = {Thorpe, J. I. and Slutsky, J. and Baker, John G. and Littenberg, Tyson B. and Hourihane, Sophie and Pagane, Nicole and Pokorny, Petr and Janches, Diego and (The LISA Pathfinder Collaboration) and Armano, M. and Audley, H. and Auger, G. and Baird, J. and Bassan, M. and Binetruy, P. and Born, M. and Bortoluzzi, D. and Brandt, N. and Caleno, M. and Cavalleri, A. and Cesarini, A. and Cruise, A. M. and Danzmann, K. and de Deus Silva, M. and De Rosa, R. and Di Fiore, L. and Diepholz, I. and Dixon, G. and Dolesi, R. and Dunbar, N. and Ferraioli, L. and Ferroni, V. and Fitzsimons, E. D. and Flatscher, R. and Freschi, M. and García Marirrodriga, C. and Gerndt, R. and Gesa, L. and Gibert, F. and Giardini, D. and Giusteri, R. and Grado, A. and Grimani, C. and Grzymisch, J. and Harrison, I. and Heinzel, G. and Hewitson, M. and Hollington, D. and Hoyland, D. and Hueller, M. and Inchauspé, H. and Jennrich, O. and Jetzer, P. and Johlander, B. and Karnesis, N. and Kaune, B. and Korsakova, N. and Killow, C. J. and Lobo, J. A. and Lloro, I. and Liu, L. and López-Zaragoza, J. P. and Maarschalkerweerd, R. and Mance, D. and Martín, V. and Martin-Polo, L. and Martino, J. and Martin-Porqueras, F. and Madden, S. and Mateos, I. and McNamara, P. W. and Mendes, J. and Mendes, L. and Nofrarias, M. and Paczkowski, S. and Perreur-Lloyd, M. and Petiteau, A. and Pivato, P. and Plagnol, E. and Prat, P. and Ragnit, U. and Ramos-Castro, J. and Reiche, J. and Robertson, D. I. and Rozemeijer, H. and Rivas, F. and Russano, G. and Sarra, P. and Schleicher, A. and Shaul, D. and Sopuerta, C. F. and Stanga, R. and Sumner, T. and Texier, D. and Trenkel, C. and Tröbs, M. and Vetrugno, D. and Vitale, S. and Wanner, G. and Ward, H. and Wass, P. and Wealthy, D. and Weber, W. J. and Wissel, L. and Wittchen, A. and Zambotti, A. and Zanoni, C. and Ziegler, T. and Zweifel, P. and (The ST7-DRS Operations Team) and Barela, P. and Cutler, C. and Demmons, N. and Dunn, C. and Girard, M. and Hsu, O. and Javidnia, S. and Li, I. and Maghami, P. and Marrese-Reading, C. and Mehta, J. and O’Donnell, J. and Romero-Wolf, A. and Ziemer, J.},
title = {Micrometeoroid Events in LISA Pathfinder},
journal = {The Astrophysical Journal}
}

@article{Rigley10.1093/mnras/stab3482,
    author = {Rigley, Jessica K and Wyatt, Mark C},
    title = {Comet fragmentation as a source of the zodiacal cloud},
    journal = {Monthly Notices of the Royal Astronomical Society},
    volume = {510},
    number = {1},
    pages = {834-857},
    year = {2021},
    month = {12},
    issn = {0035-8711},
    doi = {10.1093/mnras/stab3482},
    url = {https://doi.org/10.1093/mnras/stab3482},
    eprint = {https://academic.oup.com/mnras/article-pdf/510/1/834/41858148/stab3482.pdf},
}

@ARTICLE{Stark2009ApJ...707..543S,
       author = {{Stark}, Christopher C. and {Kuchner}, Marc J.},
        title = "{A New Algorithm for Self-consistent Three-dimensional Modeling of Collisions in Dusty Debris Disks}",
      journal = {\apj},
         year = 2009,
        month = dec,
       volume = {707},
       number = {1},
        pages = {543-553},
          doi = {10.1088/0004-637X/707/1/543},
archivePrefix = {arXiv},
       eprint = {0909.2227},
 primaryClass = {astro-ph.SR},
       adsurl = {https://ui.adsabs.harvard.edu/abs/2009ApJ...707..543S}
}
\bibliographystyle{aasjournalv7}
\end{document}